\documentclass{CVM}

\usepackage{wrapfig}
\graphicspath{{./figures/}}

\newcommand{\mathdash}[1]{{\operatorname{#1}}}

\usepackage{adjustbox}

\usepackage{hyperref}
\hypersetup{
    colorlinks=true,
    linkcolor=blue,
    filecolor=cyan,      
    urlcolor=magenta,
}
\usepackage{url}

\usepackage{textcomp}
\usepackage[nameinlink]{cleveref}
\usepackage{orcidlink}

\usepackage{colortbl}

\CVMsetup{
    type      = {Review Article},
    doi       = {xxxxxx-xxx-xxxx-x},
    title     = {Enhancing Efficiency in Vision Transformer Networks: Design Techniques and Insights},
    author    = {Moein~Heidari*$^{1}$, Reza~Azad*$^{2}$, Sina~Ghorbani~Kolahi*$^{3}$, René~Arimond$^{2}$, Leon~Niggemeier$^{2}$, Alaa~Sulaiman$^{4}$, Afshin~Bozorgpour$^{5}$, Ehsan~Khodapanah~Aghdam$^{6}$, Amirhossein~Kazerouni$^{7}$,
    Ilker~Hacihaliloglu$^{8}$,  and~Dorit~Merhof$^{5,9}$\cor{}},
    runauthor = {M.~Heidari, R.~Azad, S.~G.~Kolahi, \textit{et al.}},
    abstract  = {Intrigued by the inherent ability of the human visual system to identify salient regions in complex scenes, attention mechanisms have been seamlessly integrated into various Computer Vision (CV) tasks. Building upon this paradigm, Vision Transformer (ViT) networks exploit attention mechanisms for improved efficiency. This review navigates the landscape of redesigned attention mechanisms within ViTs, aiming to enhance their performance. This paper provides a comprehensive exploration of techniques and insights for designing attention mechanisms, systematically reviewing recent literature in the field of CV. This survey begins with an introduction to the theoretical foundations and fundamental concepts underlying attention mechanisms. We then present a systematic taxonomy of various attention mechanisms within ViTs, employing redesigned approaches. A multi-perspective categorization is proposed based on their application, objectives, and the type of attention applied. The analysis includes an exploration of the novelty, strengths, weaknesses, and an in-depth evaluation of the different proposed strategies. This culminates in the development of taxonomies that highlight key properties and contributions. Finally, we gather the reviewed studies along with their available open-source implementations at our \href{https://github.com/mindflow-institue/Awesome-Attention-Mechanism-in-Medical-Imaging}{GitHub}\footnote{\url{https://github.com/xmindflow/Awesome-Attention-Mechanism-in-Medical-Imaging}}. We aim to regularly update it with the most recent relevant papers.

 },
    keywords  = {Attention mechanisms, computer vision, deep learning, vision transformer (ViT), transformer},
    copyright = {The Author(s) \the\year{}.},
}

\begin{document}

\maketitle

\begin{figure}[!b] \vskip -4mm
\small\renewcommand\arraystretch{1.3}
    \begin{tabular}{p{80.5mm}} \toprule\\ \end{tabular}
    \vskip -4.5mm \noindent \setlength{\tabcolsep}{1pt}
    \begin{tabular}{p{3.5mm}p{80mm}}
        $1\quad $ & School of Biomedical Engineering, University of British Columbia, British Columbia, Canada. E-mail: \texttt{moein.heidari@ubc.ca}.
        \\
        $2\quad $ & Faculty of Electrical Engineering and Information Technology, RWTH Aachen University, Aachen, Germany. E-mail: \texttt{\{reza.azad; rene.arimond; leon.niggemeier\}@rwth-aachen.de}.
        \\
        $3\quad $ & Department of Industrial and Systems Engineering, Tarbiat Modares University, Tehran, Iran. E-mail: \texttt{sina.ghorbani@modares.ac.ir}.
        \\
        $4\quad $ & Faculty of Information Science and Technology, Universiti Kebangsaan, Bangi, Malaysia. E-mail: \texttt{alaasol@gmail.com}.
        \\
        $5\quad $ & Faculty of Informatics and Data Science, University of Regensburg, Regensburg, Germany. E-mail: \texttt{\{afshin.bozorgpour; dorit.merhof\}@ur.de}.
        \\
        $6\quad $ & Department of Electrical Engineering, Shahid Beheshti University, Tehran, Iran. E-mail: \texttt{ehsan.khpaghdam@gmail.com}.
        \\
        $7\quad$ & Department of Computer Science, University of Toronto, Toronto, Canada. E-mail: \texttt{amirhossein@cs.toronto.edu}
        \\
        $8\quad $ & Department of Radiology, Department of Medicine, University of British Columbia, British Columbia, Canada. E-mail: \texttt{ilker.hacihaliloglu@ubc.ca}
        \\
        $9\quad $ & Fraunhofer Institute for Digital Medicine MEVIS, Bremen, Germany.
        \\
        * & \textit{Indicates equal contribution}
        \\
        &\hspace{-5mm} Manuscript received: 2024-01-01; accepted: 2024-04-04\vspace{-2mm}
    \end{tabular} \vspace{-3mm}
\end{figure}

\section{Introduction} \label{sec:intro}

Attention mechanisms help the human visual system to efficiently and effectively analyze and comprehend complex scenes~\cite{itti1998model} by focusing on the essential areas of an image while ignoring irrelevant parts. Inspired by this concept, attention mechanisms have been introduced in Computer Vision (CV) to dynamically assign weights to different regions within an image. This enables neural networks to focus on significant areas relevant to the target task while ignoring unimportant regions.
Following their influential introduction in natural language processing (NLP)~\cite{bahdanau2014neural} to overcome the drawbacks of traditional neural networks, attention mechanisms have achieved immense success in diverse tasks. Notably, they have been effectively utilized in various tasks, including text classification~\cite{Liu2019bidirectional,li2019improving}, image segmentation~\cite{dosovitskiy2020vit,chen2021transunet,liu2021swin,chen2021crossvit,huang2022missformer,zhou2021latent,chen2017sca,xu2015show,fiaz2022sa2}, machine translation~\cite{sutskever2014sequence,luong2015effective,britz2017massive,vaswani2017attention} and speech recognition~\cite{chorowski2014end,chan2016listen,sperber2018self}~\cite{niu2021review}.

The powerful capabilities of attention mechanisms are well suited for capturing complex semantic relationships in visual data. In CV, objects of interest are often confined to small regions and appear at different scales within the input, posing challenges for conventional architectures. Attention networks are therefore used to alleviate these problems by forcing the model to focus on informative locations while ignoring non-informative ones. Recently, considerable research in CV has focused on deep neural structures known as Vision Transformers (ViT)~\cite{dosovitskiy2020vit}, which rely on self-attention mechanisms. However, standard self-attention as used in ViTs suffers from quadratic computational and memory complexity, limiting its ability to process high-resolution inputs and scale to downstream tasks. This has motivated considerable research into modifications such as sparse attention patterns~\cite{wang2021pyramid, wang2022pvt}, constrained local contexts~\cite{pan2023slide, liu2021swin,  liu2021swinv2} and efficient attention mechanisms \cite{huang2022missformer, shen2021efficient, ali2021xcit}. In addition to the advancements in self-attention mechanisms in ViTs, it is important to emphasize that the design of transformers for CV requires an adaptive strategy to capture hierarchical feature descriptions \cite{heidari2022hiformer}. This adaptation is necessary because objects of interest in visual data often have different shapes and scales, requiring a flexible approach to accurately represent and analyze the variety of visual patterns encountered. Moreover, the tokenization process in ViTs plays a pivotal role in improving computational efficiency. Careful consideration and optimization of tokenization methods (e.g., resampling techniques~\cite{xia2022vision, zhu2023biformer}) contribute significantly to the overall performance of ViT models. Efficient tokenization not only facilitates better computation, but also improves the model's efficiency in handling diverse input data.

Furthermore, it is noteworthy that addressing the challenges associated with self-attention in ViTs involves exploring diverse attention mechanisms, including spatial and channel attention \cite{ding2022davit}. These modifications aim to improve computational efficiency while maintaining performance. In summary, enhancing the structure of the ViT is crucial to enable efficient and scalable attention mechanisms in CV. Considerable research efforts have been devoted to exploring the utility of attention for CV, resulting in a substantial influx of contributions in this burgeoning field. Consequently, a survey of the existing literature is not only beneficial but also timely for the community. With this goal in mind, this review aims to provide a comprehensive overview of recent advances and to presents a holistic view of attention-based models for CV. We characterize technical innovations and major use cases through proposed taxonomies, examine the background of attention in vision, and elaborate on well-known architectures such as transformers. We review key technologies that have emerged from various CV applications, including image segmentation, registration, reconstruction, and classification. The intention of our work is to identify novel research opportunities, provide guidance, and stimulate interest in the use of attention networks for CV.

The specific contributions of this paper can be summarized as follows:

\begin{itemize}

\item We systematically and comprehensively review the design and intuition behind the attention mechanism by proposing a unified model. This includes respective taxonomies, and discussions of various aspects of the attention mechanism. 


\item Our objective is to meticulously and systematically examine the range of attention mechanisms integrated within the transformer network, all directed at optimizing its efficiency. We divide the existing research into four categories (\Cref{fig-taxonomy-design}): \textbf{\textit{Self-Attention Complexity Reduction}}, \textbf{\textit{Hierarchical Transformer}}, \textbf{\textit{Channel and Spatial Transformer}}, \textbf{\textit{Rethinking Tokenization}}, and \textbf{\textit{Other}}.  This categorization provides a systematic overview of different design techniques for attention mechanisms in CV, particularly within ViTs. The exploration also encompasses contributions to transformer architectures for various CV tasks. 


\item Finally, we discuss challenges and open issues, and identify emerging trends, open research questions, and future directions in the context of enhanced ViTs.

\end{itemize}

\subsection{Search Strategy}

We conducted a thorough search using DBLP, Google Scholar, and Arxiv Sanity Preserver, using customized search queries that allowed us to obtain lists of scientific publications. These publications included peer-reviewed journal papers, conference or workshop papers, non-peer-reviewed papers, and preprints. Our search queries consisted of the keywords \texttt{(attention* $|$ deep* $|$ efficient*), (transformer $|$ efficient*), (transformer* $|$ efficient* $|$ image* $|$ attention*), (attention* $|$ vision* $|$ transformer* $|$ medical*)}. To ensure the selection of relevant papers, we carefully evaluated their novelty, contribution, and significance, and prioritized those that were the first of their kind in the field of CV. Following these criteria, we selected papers with the highest rankings for further consideration. It is worth noting that our review may have excluded other significant papers in the field, but our goal was to provide a comprehensive overview of the most important and impactful papers.

\subsection{Paper Organization}

The paper is organized as follows. In \Cref{sec:background}, we provide a detailed overview of the concepts and theoretical foundations underlying attention models. We elucidate these concepts by introducing two taxonomies and a unified attention model. \Cref{chap:transformer-intro} delves into the intricate structure of the ViT architecture. The focus of our work is captured in \Cref{chap:attention-design}, where we present a taxonomy that categorizes efficient attention mechanism designs specifically within ViTs. Sections \ref{SA-reduce} to \ref{other} comprehensively review the methods outlined in \Cref{fig-taxonomy-design}, and \Cref{sec:discuss} provides a comprehensive discussion of the approaches presented in this work. Next, \Cref{sec:challenge} outlines open challenges and future perspectives for the field as a whole. Finally, \Cref{analysis-sec} conducts an in-depth analysis of ViTs attention blocks based on the proposed taxonomy.


\subsection{Motivation and Uniqueness of Survey}
The recent surge of interest surrounding the exceptional performance of the transformer architecture in NLP has been seamlessly transitioned to the field of CV~\cite{vaswani2017attention,dosovitskiy2020vit}. Renowned for their proficiency in capturing long-range dependencies and spatial correlations through their attention-centric nature, transformers present a clear advantage over the conventional convolutional neural networks (CNNs) that have historically dominated CV tasks. While numerous survey papers have explored attention mechanisms and ViT models, existing works often narrow their focus to specific applications or modalities~\cite{azad2023advances,azad2023foundational, xu2023multimodal,ulhaq2022vision, VideoTransformers}. For instance, Brauwers et al.~\cite{AttentionSUrvey} provide a general explanation of attention and an overview of attention techniques in deep learning, regardless of data modality. Similarly, Guo et al.~\cite{guo2022attention} provide a comprehensive review of various attention mechanisms in CV, categorizing them according to approach. In contrast to these, some surveys focus on the evolution of visual transformers specific to CV tasks~\cite{Khan_2022,Han_2023, VisualTransformers} Furthermore, in the context of efficient ViTs, Patro et al.~\cite{patro2023efficiency} provide a comprehensive compilation of efficient variants, categorized according to factors such as computational complexity, robustness, and transparency. Nauen et al.~\cite{nauen2023transformer} examine the efficiency of ViTs and their architectural modifications, focusing on parameters, FLOPs, speed, and memory during training on ImageNet1k~\cite{imagenet}. 

Our paper distinguishes itself by presenting a comprehensive investigation of the general form of attention mechanisms and their applications in CV. We revisit the ViT architecture and provide a comprehensive and up-to-date review of recent efficient ViT models. 
Significantly, we introduce a novel taxonomy designed to categorize and enhance ViT networks based on their attention mechanisms and approaches, beyond the constraints of specific CV tasks (see~\Cref{fig-taxonomy-design}). Our review also includes real-world applications of efficient ViTs. Leveraging our proposed taxonomy, we conduct an in-depth analysis of ViT attention blocks, comparing their advantages and drawbacks based on contributions and numerical metrics such as the number of parameters, FLOPS (Floating Point Operations), MACs (Multiply-Accumulate Operations), and time complexity (\Cref{sec:discuss}). We also explore the challenges and future directions of this emerging field. This approach distinguishes our work, providing a unique perspective and contribution to the understanding and optimization of ViT models in the context of CV.

\subsection{Real-World applications}

In recent years, transformer models and their enhanced variants have reshaped the landscape of CV, demonstrating remarkable success in core tasks such as image recognition ~\cite{dosovitskiy2020vit,liu2021swin, fan2021multiscale, li2022mvitv2, liu2021swinv2,touvron2021training}, object detection ~\cite{carion2020end,maaz2022class,meng2021conditional,zhang2022dino,zhu2020deformable}, and segmentation ~\cite{strudel2021segmenter,xie2021segformer,cheng2022masked,shaker2022unetr++}. Their adaptability extends to more complex-level CV challenges, including video analysis ~\cite{sun2019videobert, arnab2021vivit}, image/video generation ~\cite{chen2021pre, zhang2021styleswin,chen2023pixartalpha}, super resolution ~\cite{zou2022self,lu2022transformer,zhang2022efficient,geng2022rstt}, real-time mobile vision ~\cite{zhou2022crossview, wang2022rtformer}. The efficiency gains achieved through enhanced ViTs result in substantial reductions in training and inference times, making them pivotal in real-time scenarios. Moreover, their integration into resource-constrained environments, such as mobile devices, not only extends advanced vision capabilities to a broader user base but also reduces deployment costs. This adaptability aligns with the broader push for environmentally sustainable practices in AI, as the streamlined architectures contribute to lower carbon footprints during model training.

Furthermore, the transformative impact of efficient ViTs is evident in critical domains like healthcare. The high precision and adaptability of these models facilitate the development of advanced tools for Clinical Decision Support Systems ~\cite{long2021edssr,bai2023catvil}. Empowering healthcare professionals with more accurate and timely insights, these tools contribute to improved diagnostic capabilities and patient outcomes. As enhanced ViTs continue to evolve, their versatility and high performance position them as indispensable solutions for addressing a diverse array of real-world vision challenges ~\cite{kinfu2023efficient,Song_2022_CVPR}.


In the field of image/video super-resolution, researchers have been actively exploring innovative approaches to enhance the capabilities of ViTs in practical applications. Geng et al.~\cite{geng2022rstt} propose a Real-Time Spatial Temporal Transformer (RSTT) for Space-Time Video Super-Resolution (STVSR). This transformer integrates temporal interpolation and spatial super-resolution modules into a unified framework, resulting in a more compact network compared to existing methods. The RSTT achieves real-time inference speed without significant performance loss. Notably, the authors present the RSTT as the first application of a transformer to address the STVSR problem. Within the RSTT, a cascaded UNet-style architecture effectively integrates spatial and temporal information for synthesizing High Frame Rate (HFR) and High-Resolution (HR) video. The encoder part of the RSTT builds multi-resolution dictionaries, which are then queried in the decoder part for directly reconstructing HFR and HR frames. Experimental results demonstrate that the RSTT is significantly smaller and faster than state-of-the-art STVSR methods while maintaining similar performance levels.

Researchers have also turned their attention to addressing the challenges of real-time mobile vision tasks. This expanding domain requires unique solutions that meet the demands for speed and efficiency in processing visual information on mobile devices, while also incorporating eco-friendly approaches to develop them. In their innovative work, Wang et al.~\cite{wang2022rtformer} present the RTFormer block, an efficiently designed transformer for GPU-like devices, with a focus on achieving an optimal balance between performance and efficiency. Introducing GPU-Friendly Attention in the low-resolution branch addresses multi-head mechanism limitations, ensuring linear complexity and improved parameter utilization. The high-resolution branch incorporates cross-resolution attention and a stepped layout, enhancing the integration of global context information from the low-resolution branch. This innovative RTFormer block is employed to construct the RTFormer real-time semantic segmentation network, strategically positioned in the last two stages. Through extensive experiments, the study demonstrates that RTFormer attains a more refined balance between performance and efficiency when compared to previous methodologies. 

PIXART-$\alpha$~\cite{chen2023pixartalpha} addresses the substantial training costs associated with advanced text-to-image (T2I) models, impeding innovation and contributing to increased CO2 emissions. PIXART-$\alpha$, the proposed transformer-based T2I diffusion model, achieves competitive image generation quality comparable to state-of-the-art generators (e.g., Imagen~\cite{saharia2022photorealistic}, SDXL~\cite{podell2023sdxl}), meeting near-commercial application standards. Notably, PIXART-$\alpha$ supports high-resolution image synthesis up to 1024px with reduced training costs. The core designs include a decomposed training strategy, an efficient T2I transformer with cross-attention modules, and a focus on high-informative data. PIXART-$\alpha$'s training speed outperforms existing large-scale models, with a marked reduction in training time, saving costs, and significantly reducing CO2 emissions. This model demonstrates superiority in image quality, artistry, and semantic control through extensive experiments. The authors aim to offer valuable insights, facilitating the development of high-quality, cost-effective generative models.


Besides, ViTs are crucial in medical applications, particularly in reconstructing surgical scenes for robotic-assisted surgery, improving trainee understanding despite obstructed views~\cite{long2021edssr}. They address medical challenges by automating knowledge dissemination and providing solutions to the scarcity of expert insights. This includes innovative applications such asVisual Question Answering (VQA) models in the medical domain, ensuring efficient and comprehensive learning~\cite{bai2023catvil}.

According to this importance, Long et al.~\cite{long2021edssr} introduce E-DSSR, an Efficient Dynamic Surgical Scene Reconstruction pipeline, enhancing stereoscopic depth perception exclusively from stereo endoscopic images. It improves upon prior works with an image-only reconstruction pipeline, incorporating a transformer-based depth perception module and a lightweight tool segment. These modules run in parallel and provide a masked depth estimation without surgical instruments. E-DSSR simultaneously handles challenges such as tissue deformation, tool occlusion, and camera movement. The results demonstrate the effectiveness of the proposed approach.

Bai et al.~\cite{bai2023catvil} propose CAT-ViL DeiT, a specialized transformer model for Visual Question Localized-Answering (VQLA) in surgical scenes, which seamlessly integrates tasks and demonstrates the potential of AI in surgical training. The CAT-ViL embedding, with its co-attention and gated modules, excels in promoting instructive text-visual interactions. With its exceptional performance, CAT-ViL DeiT efficiently locates and answers questions in surgical scenarios, outperforming alternatives in real-time applications.

Overall, a wealth of research has focused on enhancing the transformer model and its central attention mechanism to effectively adapt them for practical use in various real-world scenarios.

\section{Background}\label{sec:background}

In this chapter, we define the necessary background of the attention mechanism and the scope of this survey. First, the attention mechanism is introduced~\cite{guo2022attention} and a unified attention model~\cite{niu2021review} is presented. Then, two taxonomies to categorize attention mechanisms are shown.
Lastly, the transformer~\cite{vaswani2017attention} and ViT~\cite{dosovitskiy2020vit} architectures are explained, including the underlying attention mechanism.

Building upon this understanding, the chapter proceeds to elucidate the most influential architectures in the field of CV these days - the ViT networks~\cite{dosovitskiy2020vit}.

\subsection{Attention}
The attention mechanism is a fundamental cognitive process that humans utilize daily to navigate their surroundings effectively. It plays a crucial role in determining \textit{what}, \textit{when}, and \textit{where} individuals choose to direct their cognitive resources. This selectivity ensures that humans do not become overwhelmed by an excessive influx of sensory information, such as visual, auditory, or tactile stimuli, but rather focus on what is most relevant and significant at any given moment. By prioritizing specific information, the attention mechanism optimizes the accuracy and performance of human information processing, allowing individuals to efficiently interact with the world around them~\cite{guo2022attention}.

Human attention can be broadly categorized into two main types: \textit{unfocused} and \textit{focused} attention. Unfocused attention operates automatically and involuntarily, meaning it cannot be actively influenced by conscious decisions. Instead, it operates as a background process that continuously monitors the environment for potential salient cues or changes, without conscious control from the individual. On the other hand, focused attention allows humans to deliberately and consciously direct their cognitive spotlight onto a particular object, task, or aspect of their surroundings. This focused attentional control enables humans to engage in complex and demanding cognitive tasks effectively~\cite{heitz2007focusing}.

Interestingly, the attention mechanism in deep learning models exhibits a parallel with human-focused attention in many cases. In deep neural networks, the attention mechanism serves the critical purpose of allocating resources to the most relevant and informative parts of the input data. By doing so, it empowers machines to efficiently tackle complex visual or language tasks, even when computational resources are limited. Similar to human-focused attention, the attention mechanism in deep learning enables the model to focus on crucial aspects of the task at hand, facilitating more accurate and meaningful outcomes~\cite{niu2021review}.

When applied to visual tasks, such as object detection or image captioning, the attention mechanism allows the model to selectively attend to specific regions of an image, emphasizing vital features and downplaying irrelevant ones~\cite{guo2022attention}. Likewise, in NLP tasks, the attention mechanism enables the model to emphasize the most important words or phrases in a sentence or document, capturing the context and semantics effectively~\cite{AttentionSUrvey}. By employing attention, deep neural networks can leverage the power of focused processing, just as humans do when addressing complex cognitive challenges.

The attention mechanism has emerged as a powerful tool in deep learning era, finding successful applications across a wide range of tasks. 
In the next subsection, we will thoroughly examine the definition and workings of the unified attention model. Then we will introduce the ViT model.

\subsection{The General Attention Mechanism}
\begin{figure}[!ht]
    \centering
    \begin{subfigure}{0.5\textwidth}
        \centering
        \includegraphics[width=\textwidth]{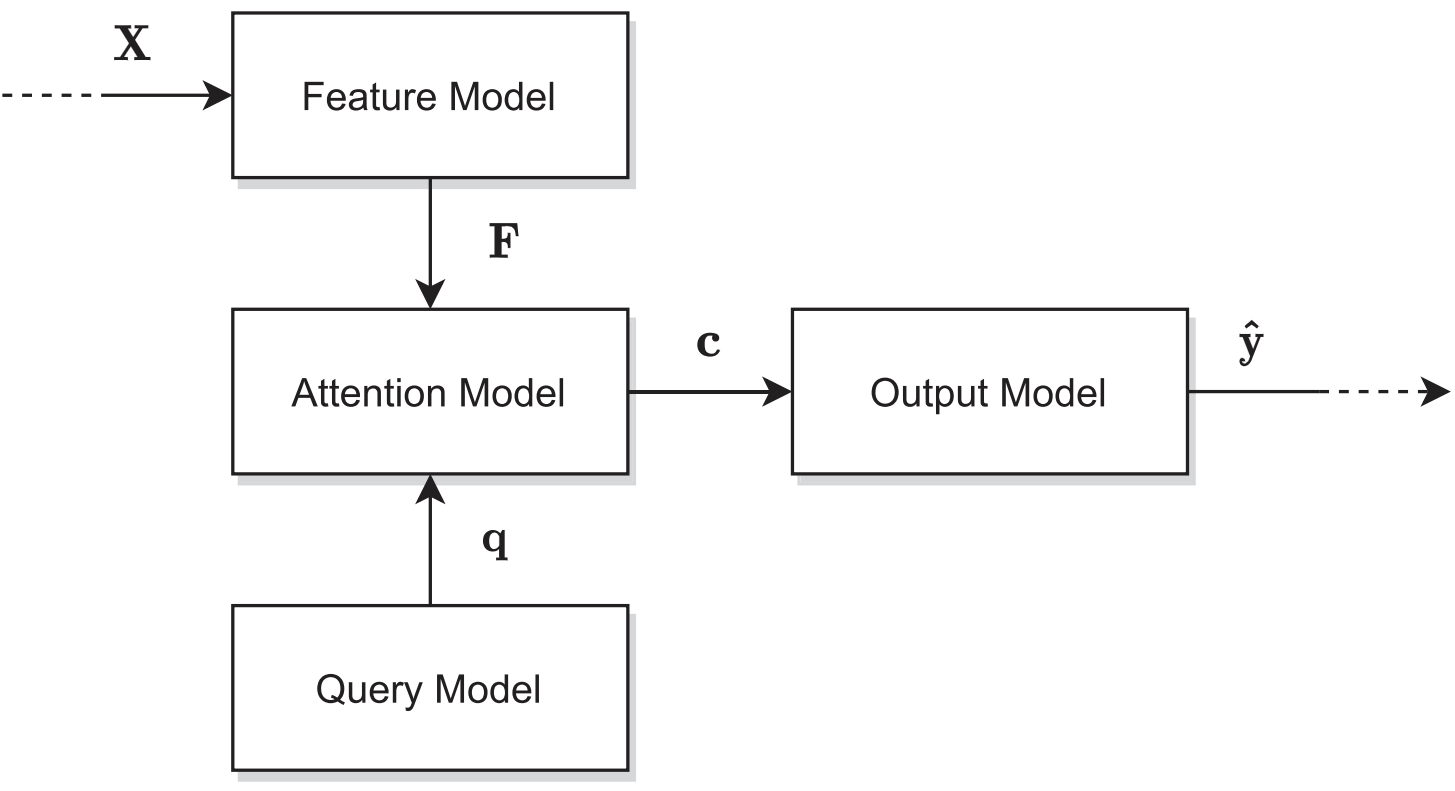}
        \caption{}
        \label{fig:taskmodel}
    \end{subfigure}
    \hfill
    \begin{subfigure}{0.45\textwidth}
        \centering
        \includegraphics[width=\textwidth]{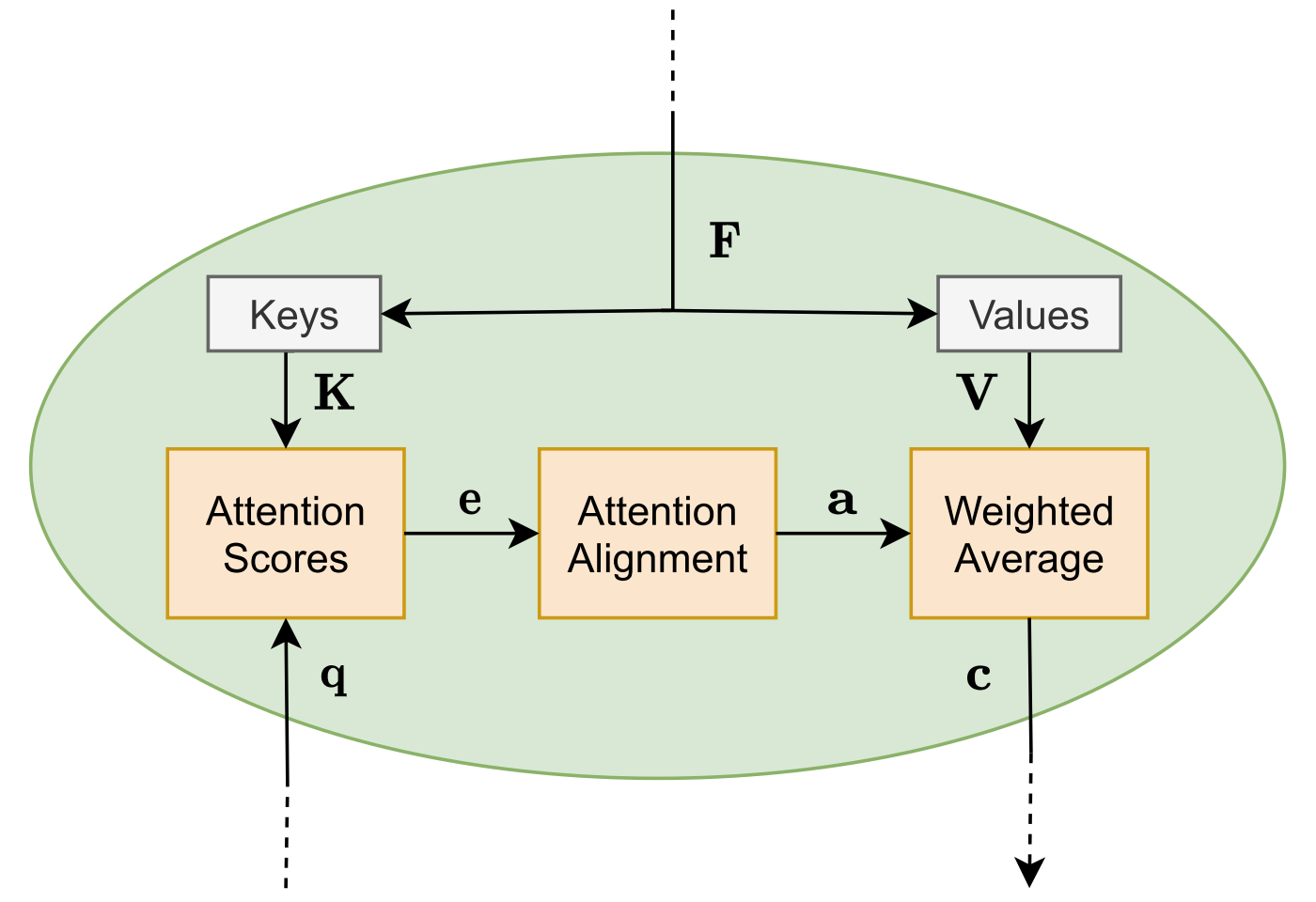}
        \caption{}
        \label{fig:generalizedattention}
    \end{subfigure}

    \label{fig:task_generalized_attention}
    \caption[Task model and generalized attention]{(a) The task model and (b) The generalized attention module. From \cite{AttentionSUrvey}.}
\end{figure}

Human attention's significance extends to CV, where attention mechanisms were introduced to address computational costs~\cite{Attentionfirst, attentionsecond}. This involves focusing on vital regions in an image, effectively reducing the processing load. The recognition of attention's importance surged after Vaswani et al.'s groundbreaking results in NLP tasks \cite{vaswani2017attention}. Various forms of attention mechanisms have since emerged, with Brauwers et al. \cite{AttentionSUrvey} introducing a comprehensive \textit{task model} illustrated in Figure \ref{fig:taskmodel}. This model takes input, performs a specific task, and produces the desired output. In applications like image segmentation, the \textit{task model's} attention mechanism proves beneficial by highlighting salient regions, thereby contributing to segmentation map precision. The \textit{task model} encompasses four sub-modules: the \textit{feature model}, the \textit{query model}, the \textit{attention model}, and the \textit{output model}.

Taking a segmentation example, the \textit{feature model} extracts distinctive features such as edges and textures from input images to facilitate precise segmentation. The \textit{query model} generates queries that guide the \textit{attention model}, prioritizing features relevant to object boundaries. The \textit{attention model}, illustrated in Figure \ref{fig:generalizedattention}, processes both feature vectors and queries, leading to the extraction of key and value matrices. A score function combines query and key matrices, resulting in attention scores. These scores, in turn, act as weighting matrices, orchestrating a weighted average of the corresponding value vectors. This strategic process helps identify key regions to ensure accurate segmentation. The \textit{output model} utilizes the attention-focused context vector to generate a segmentation map with improved precision and accuracy. Overall, this architecture optimizes the segmentation process by emphasizing key features through attention mechanisms, contributing to a more accurate final segmentation map.

\subsection{Taxonomy of Attention: A Generalized Perspective}

Our categorization is broad enough to
capture many of the models as they use fundamental ideas that are already present in
existing work and therefore can be categorized appropriately. We briefly discuss diverse preceding attention mechanism taxonomies before diving into the details of our hierarchy. 

Niu et al.~\cite{niu2021review} define attention based on four aspects:
Softness, Input Representation, Output Representation, and  Forms of Input Features. 
Each aspect is presented in the following section.


\subsubsection{Softness}

Soft attention is deterministic and uses a weighted average of all keys to build the context vector. Soft attention modules are differentiable and hence,
networks employing soft attention mechanisms can be trained by back-propagation.

Hard attention is stochastic and can be implemented as follows:

\begin{align}
	\tilde{\alpha} \sim \text{Multinoulli}(\{\alpha_{i}\}),
	\label{eq-hard-attention-1}
\end{align}
and

\begin{align}
	c = \sum_{i=1}^{n}\tilde{\alpha_{i}}\mathbf{\nu_{i}},
	\label{eq-hard-attention-2}
\end{align}
where $\mathit{Multinoulli}$ is a categorical distribution and $\tilde{\alpha_{i}} \in \tilde{\alpha}$. Hard attention makes the module less computationally expensive but disables back-propagation.

\subsubsection{Forms of Input Feature}

Input features can be distinguished as \textit{item-wise} and \textit{location-wise}. Item-wise input features
are either explicit items or a sequence of items is generated from the input.
Location-wise attention functions for tasks where the generation of explicit items is hard. In visual
tasks,~\cite{niu2021review} counts multi-resolution crops and pose transformation as location-wise attention.

\subsubsection{Input Representation}

\textit{Distinctive attention} requires a single input and output sequence. The keys and queries are sampled from different sequences.
~\textit{Co-attention} requires multiple inputs, which can be processed sequentially or parallelly, coarse-grained or fine-grained, and is used for a visual question-answering task in \cite{lu2016hierarchical}.
In \textit{self-attention}, \textbf{q}, \textbf{K}, and \textbf{V} are representations of the same input data. The transformer~\cite{vaswani2017attention} model relies on self-attention.
When using \textit{hierarchical attention}, the attention weights are not only computed from the input but also from different abstraction levels. Hierarchies
can be document-level, sentence-level, and word-level for language tasks or object-level and part-level for CV tasks.

\subsubsection{Output Representation}

The output may be \textit{single-output} - a single vector at each time step.
Another option is \textit{multi-head} attention. Here, multiple different attention weight vectors are learned and then concatenated. This principle is used in the transformer architecture~\cite{vaswani2017attention}.
Lastly, \textit{multi-dimensional} attention employs a weight score matrix instead of a vector. By doing that, each key becomes a feature-wise score vector and multiple attention distributions are computed from the same input tensor.

\subsection{Attention in Computer Vision}

Guo et al.~\cite{guo2022attention} introduce another way to classify attention modules, specifically aimed at CV tasks.
They differentiate between channel attention, spatial attention, temporal attention, and branch attention, and the two combinations of channel and spatial attention and spatial and temporal attention. 


Guo et al. introduce a simpler formula for attention:

\begin{align}
	\textnormal{Attention} = f(g(\mathbf{x}),\mathbf{x}),
	\label{eq-attention-simple}
\end{align}
where $g(\mathbf{x})$ represents the distribution function from the unified model and $f(g(\mathbf{x}),\mathbf{x})$ represents the context vector $\mathbf{c}$.

\subsubsection{Channel Attention}

Channel attention was first introduced in the SENet~\cite{SEnet}. Channel attention is a way to recalibrate channel weights - it determines what to pay attention to.
Each channel usually represents a different feature map of the same input, hence channel recalibration assigns different importance to different objects.


\subsubsection{Spatial Attention}

Spatial attention focuses on the \textit{where}. Modules employing spatial attention adaptively select regions. Examples for spatial attention modules are Non-Local~\cite{wang2018nonlocal}, RAM~\cite{mnih2014recurrent}, STN~\cite{jaderberg2015spatial}, and GENet~\cite{hu2018gather}. Non-local ~\cite{wang2018nonlocal} is a spatial self-attention module that computes the dot-product between query and key. RAM~\cite{mnih2014recurrent} uses RNN to recurrently predict important regions. STN~\cite{jaderberg2015spatial} uses a sub-network to predict an affine transformation. GENet~\cite{hu2018gather} uses average pooling to recalibrate the spatial feature. This computation captures long-range spatial context.

\subsubsection{Temporal Attention}

Temporal attention is a process to adaptively select \textit{when to pay attention}. It is mostly used for video processing,
as image data does not have a time dimension. Example approaches are GLTR~\cite{li2019gltr} and TAM~\cite{liu2021tam}.

\subsubsection{Branch Attention}

When applying branch attention, one selects \textit{which to pay attention to}. A branch refers to a conditional unit in a network that controls the information flow through
the layers. This can be implemented as a highway network~\cite{srivastava2015training}, which combines different branches. Another approach is adaptive convolution kernel selection, called CondConv~\cite{yang2019condconv}, which combines multiple convolution kernels dynamically.

\subsubsection{Channel and Spatial Attention}

Channel and spatial attention combines selecting important objects - channel attention - and important regions - spatial attention.
An example is the residual attention network~\cite{wang2017residual}, which utilizes both a trunk and a mask branch.
The mask branch reweighs the output feature of the trunk branch. This network, however, fails at learning long-distance relations. In order to rectify this issue, the CBAM~\cite{woo2018cbam} was proposed.
The CBAM, short for convolutional block attention module, sequentially combines channel and spatial attention. Other implementations of the channel and spatial attention are, among others BAM - bottleneck attention module~\cite{park2018bam},
scSE - spatial and channel SE blocks~\cite{roy2018recalibrating}, Triplet attention~\cite{misra2021rotate}, SimAM~\cite{yang2021simam},
Coordinate attention~\cite{hou2021coordinate}, DANet - dual attention network~\cite{fu2019dual}, RGA - relation-aware global attention~\cite{zhang2020relation},
Self-calibrated convolutions~\cite{liu2020improving}, SPNet - strip pooling net~\cite{hou2020strip}, SCA-CNN - spatial and channel-wise attention-based convolutional neural network~\cite{chen2017sca}
and GALA - global and local attention~\cite{linsley2018gala}.

\subsubsection{Spatial and Temporal Attention}

As the name suggests, spatial and temporal attention combines selecting important regions and keyframes in a video sequence.
This type of attention is not relevant to this paper and therefore not explored further.

\section{Transformer Networks}
\label{chap:transformer-intro}
In this section, the purpose and functionality of transformers are outlined.

\subsection{Transformer Architecture}
Vaswani et al.~\cite{vaswani2017attention} introduced a new architecture for machine translation, namely the transformer.
The main problem with previous approaches - RNNs, LSTM~\cite{hochreiter1997lstm} and GRU~\cite{chung2014gru} - is that recurrent
network architectures are inherently sequential and therefore offer no efficient way of computation. Previous recurrent models relied on an encoder-decoder structure for machine translation tasks, where
inputs are sequences of tokens $\mathbf{x} = (x_{1},\ldots,x_{n})$ that are mapped to sequences of continuous representations $\mathbf{z} = (z_{1},\ldots,z_{n})$.
From $\mathbf{z}$, the decoder generates the output sequence
symbol by symbol.

It also relies on an encoder and a decoder branch. The encoder consists of multiple identically structured layers.
Each layer is made up of two sub-layers, a multi-head attention block, and a position-wise fully connected feed-forward network. A residual connection is placed around
each sub-layer, and layer normalization concludes a sub-layer.

The decoder works similarly, also employing a stack of 6 identical layers. On top of the two sub-layers from the encoder, the decoder utilizes masked multi-head attention,
meaning that only previous positions can be seen by the attention block, and predictions at position $i$ do not know outputs at the following positions.
The transformer offers a solution that enables parallelization and also reaches state-of-the-art results.

\subsubsection{The Vision Transformer Model}

\begin{figure*}
	\centering
    \includegraphics[width=1.55\columnwidth]{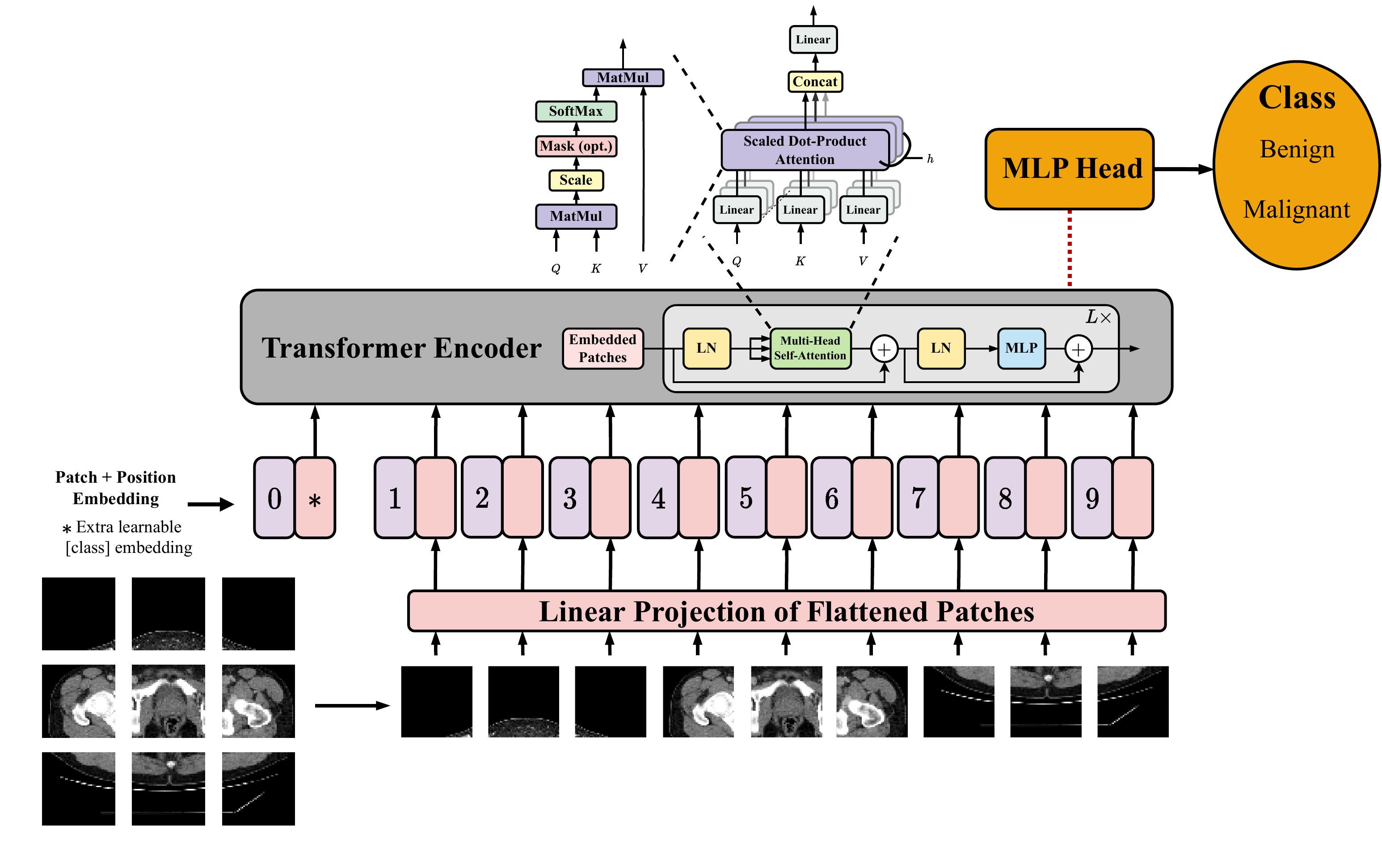}
	\caption{The Vision Transformer architecture from~\cite{dosovitskiy2020vit} is located on the center. Scaled dot-product attention and multi-head attention are on the top.}\label{fig-transformer-architecture}
\end{figure*}

Motivated by the remarkable success of transformers in NLP, Dosovitskiy et al.~\cite{dosovitskiy2020vit} introduced the ViT model, showcased in \Cref{fig-transformer-architecture}. ViT exhibits superior performance, particularly when trained on extensive datasets, outperforming the then-leading Convolutional networks. In their methodology, images undergo a transformation into fixed-size patches, subsequently flattened into vectors. These vectors undergo processing through a trainable linear projection layer, mapping them into $N$ vectors with a dimensionality of $d \times N$, where $N$ represents the number of patches. The results of this stage, termed patch embeddings, retain positional information through the addition of positional embeddings. Additionally, a trainable class embedding is incorporated into the patch embeddings before entering the transformer encoder.

The transformer encoder consists of multiple blocks, each containing a multi-head self-attention (MSA) block and an MLP block. Before entering these blocks, activations are initially normalized using LayerNorm (LN). Moreover, skip connections precede the LN, incorporating a duplicate of these activations into the corresponding MSA or MLP block outputs. Finally, an MLP block serves as a classification head, facilitating the mapping of outputs to class predictions. The self-attention mechanism emerges as a pivotal characteristic of transformer models, prompting an exploration of its core principle in the subsequent discussion.



\subsubsection{Attention in the Transformer: \textbf{Self-attention}}

In a self-attention layer (\Cref{fig-transformer-architecture} (Up-Right)), the input vector is first transformed into three separate vectors: the query vector $\mathbf{q}$, the key vector $\mathbf{k}$, and the value vector $\mathbf{v}$, all with a fixed dimension. These vectors are then organized into three different weight matrices, denoted as $W^{Q}$, $W^{K}$, and $W^{V}$. The general expressions for $Q$, $K$, and $V$ can be formulated as follows for an input $\mathbf{X}$:
\begin{align}
    \mathbf{K} &= W^K \mathbf{X}, \\
    \mathbf{Q} &= W^Q \mathbf{X}, \\
    \mathbf{V} &= W^V \mathbf{X}.
\end{align}

Here, $W^{K}$, $W^{Q}$, and $W^{V}$ refer to the learnable parameters. The scaled dot-product attention mechanism is then defined as:
\begin{equation}
    \text{Attention}(\mathbf{Q}, \mathbf{K}, \mathbf{V}) = \text{Softmax}\left(\frac{\mathbf{Q} \mathbf{K}^T}{\sqrt{d_k}}\right) \mathbf{V},
\end{equation}
where $\sqrt{d_k}$ is a scaling factor, and the SoftMax operation is applied to the generated attention weights to obtain a normalized distribution.

The concept of a multi-head self-attention (MHSA) mechanism has been introduced for capturing intricate relationships among token entities from diverse perspectives. Particularly, the MHSA block facilitates the model in simultaneously focusing on information within multiple representation subspaces, since the granularity of modeling by a single-head attention block is comparatively coarse. The MHSA procedure can be expressed as:

\begin{align}
    \text{MultiHead}(Q, K, V) &= \text{Concat}(\text{head}_1, \ldots, \text{head}_h) \cdot W^O,
\end{align}

where $\text{head}_i = \text{Attention}(Q \cdot W_i^Q, K \cdot W_i^K, V \cdot W_i^V)$, and $W^O$ represents a linear transformation for aggregating multi-head representations. Note that the hyper-parameter $h$ is defined as $h = 8$ in the original reference.

\subsection{Preliminary}\label{chap-related-work}

The transformer architecture has become the standard for NLP tasks. With the introduction of the ViT~\cite{dosovitskiy2020vit}, CNN-based approaches are challenged in CV tasks due to the attention mechanism's
ability to model long-range context. The standard self-attention has one major drawback, however: Its computational complexity is quadratic with respect to the number of tokens $N$. Since $N$ increases drastically for higher-resolution images, it is
not applicable unless some changes are implemented.
Multiple approaches exist to tackle this problem.

In the next chapter, several transformer architectures are presented. First, a new taxonomy is introduced that categorizes these
architectures by their design. Afterwards, multiple transformer networks are shown and their attention modules are observed in detail.
A comparison of the performance and requirements of each network is displayed. Lastly, the benefits and drawbacks of the presented methods are discussed with regard to the goal of this work.

\section{Attention Based on Design}
\label{chap:attention-design}
In this section, we present a taxonomy categorizing transformer networks by design (Figure \ref{fig-taxonomy-design}). The taxonomy comprises several categories: ``Self Attention Complexity Reduction," which aims to lower self-attention computational load through techniques like windowing and reordering; ``Hierarchical Transformer," utilizes multi-scale feature representations to enhance image comprehension and minimize computational expenses; ``Channel and Spatial Transformer," using transposed output tensors and channel attention for global context recovery; ``Rethinking Tokenization," exploring token-based modifications; and ``Other," encompassing diverse strategies like focal modulation, convolution integration, and deformable attention. This taxonomy offers a structured insight into diverse attention mechanisms' roles within CV.

\subsection{Self Attention Complexity Reduction}

Many approaches exist to directly reduce the computational complexity of the self-attention mechanism. They either reduce the number of tokens 
~\cite{huang2022missformer,hassani2023neighborhood,dong2022cswin,jiao2023dilateformer,wang2022kvt,wu2022p2t,wu2021pale,tang2022quadtree,liu2022dynamic,hassani2023dilated,wei2023sparsifiner}, shift the calculation to the channel dimension~\cite{ali2021xcit,maaz2022edgenext,bolya2022hydra} and change the order of multiplying or adding of query, key and value~\cite{shen2021efficient,chen2021crossvit,liu2023efficientvit,shaker2023swiftformer,han2023flatten,lan2021couplformerrethinking,li2023rethinking,you2023castlingvit,zhang2023fcaformer}.

\subsection{Hierarchical Transformer}
Hierarchical Vision Transformers exploit multi-scale feature representations to optimize image understanding and reduce the computational cost. Examples include~\cite{liu2021swin, wang2022pvt,chen2021regionvit,yu2022metaformer,liu2021swinv2,yan2023lawin,zhou2023nnformer, vasu2023fastvit,Xu_2023_FDViT,hatamizadeh2023global, hatamizadeh2023fastervit,pan2023slide}.

\subsection{Channel and Spatial Transformer}

To regain global context after patch merging and windowed self-attention,~\cite{ding2022davit} transpose the output tensor and also compute channel attention on it. Other architectures that apply this method are~\cite{shaker2022unetr++, lv2022scvit,sun2022fusing,huang2022channelized,ma2022knowing}.

\subsection{Rethinking of Tokenization}

Some transformer Architectures either add more tokens that carry additional information~\cite{fang2022msg,guo2022cmt,zhang2023vsa}, reduce the number of redundant tokens~\cite{jiang2021all, xu2022evo,li2023bvit,chen2023sparsevit, yang2021lite, rao2022hornet, pan2023fast, zhang2023quad, Ren_2023_ICCV,zhou2023token,tu2022maxvit}~or change the token meaning~\cite{ zhang2021token,wang2022shift,zhu2023biformer}. These fall under the category of \textit{Rethinking Tokenization}.

\subsection{Other}

Other approaches that do not belong in either of the previous categories are collected here~\cite{valanarasu2021medical,feng2022evit,yu2022unest,Guo2023,zhou2022spikformer,fan2023lightweight,wang2023crossformer,patro2023spectformer,lai2023hybrid,azad2023laplacianformer, azad2023selfattention}. Focal modulation~\cite{yang2022focal} belongs in this category, as it also extracts values and a query, but instead of calculating a matrix multiplication between a query and key,
a set of CNNs is applied to hierarchically contextualize the value while the query is unchanged. DeepViT~\cite{zhou2021deepvit} designs an attention mechanism for deeper networks,~\cite{graham2021levit,wu2021cvt} include convolutions in a transformer network, and~\cite{xia2022vision} proposes deformable attention.

\section{Transformer Architectures that Apply Self-Attention Complexity Reduction} \label{SA-reduce}

In this section, transformer architectures that apply some form of complexity reduction to the attention mechanism are presented.

\subsection{Efficient Attention}

\begin{figure}
	\includegraphics[width=\linewidth]{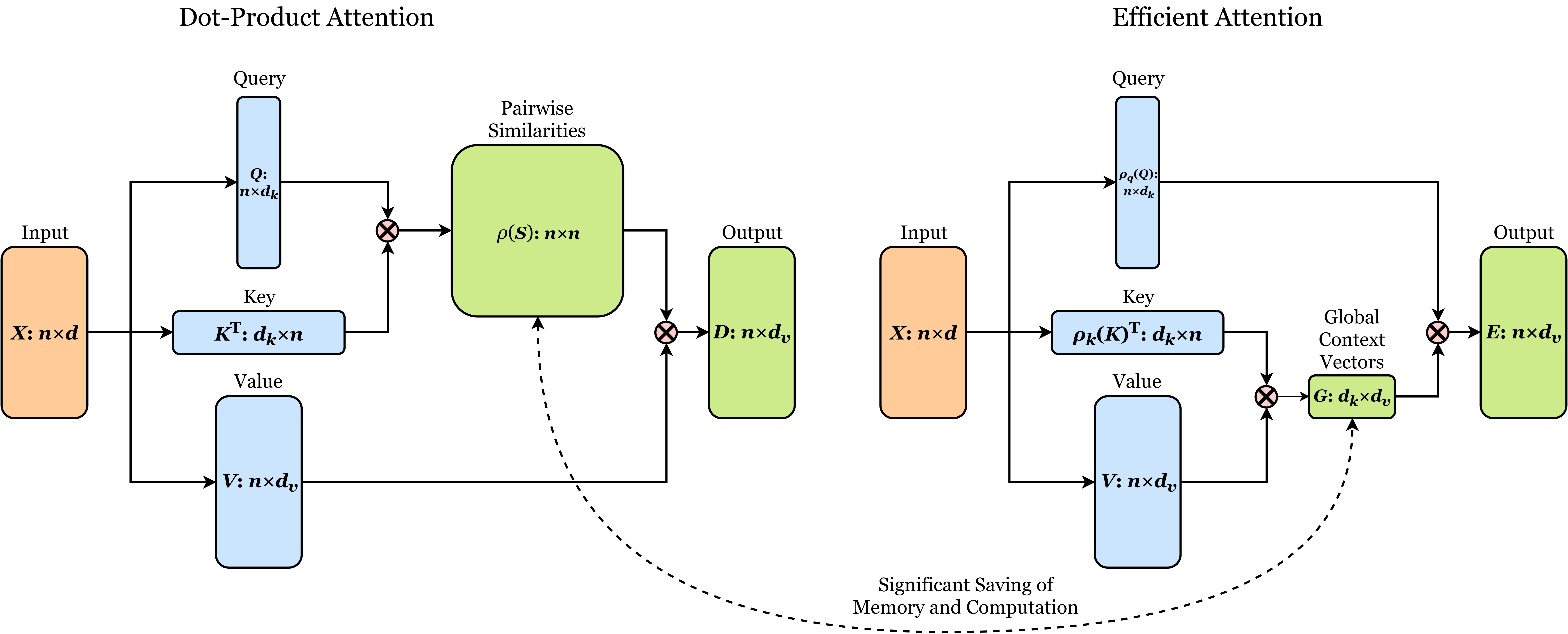}
	\caption{Standard dot-product attention on the left and efficient attention on the right. From~\cite{shen2021efficient}.}\label{fig-efficient-attention}
\end{figure}

Efficient attention, published by Shen et al.~\cite{shen2021efficient}, renews the view on the attention mechanism by shifting the order of operations.
The comparison between standard dot-product attention and efficient attention is shown in \Cref{fig-efficient-attention}.
$\rho_{q}$ and $\rho_{k}$ are normalization functions for the queries and keys. $n$ is the input size, $d$ the embedding dimension, $d_{k}$ and $d_{v}$ are the embedding dimensions of the keys and values.
When $\rho_{q}$ and $\rho_{k}$ are scaling normalization, it is proven in~\cite{shen2021efficient} that the module produces the
equivalent output of dot-product attention. When they are softmax normalization, the outputs are approximately equivalent.

Dot-product attention multiplies the queries and keys followed by a normalization step to obtain pairwise similarities. These have a dimension of
$n \times n$, with $n$ being the input dimension, whereas $d$ is the embedding dimension. Efficient attention normalizes the keys and queries first, then
multiplies the keys and values, and lastly, the resulting global context vectors are multiplied by the queries:

\begin{align}
	\mathbf{E}(\mathbf{Q},\mathbf{K},\mathbf{V}) = \rho_{q}(\mathbf{Q})(\rho_{k}(\mathbf{K})^{\mathbf{T}}\mathbf{V}).
\end{align}
Efficient attention does not, like dot-product attention, compute pairwise similarities between points first. Instead, ``it interprets the keys [\ldots] as $d_{k}$ attention maps $\mathbf{k^{T}}_{j}$"~\cite{shen2021efficient}.
These global attention maps represent a semantic aspect of the whole input feature instead of similarities to the position of the input. This shifting of orders drastically reduces the computational complexity of the attention mechanism while maintaining a high representational power.
The memory complexity of efficient attention is $O(dn + d^{2})$
while the computational complexity is $O(d^{2}n)$ when $d_{v} = d, d_{k} = \frac{d}{2}$, which is a typical setting.

\begin{figure*}[!thb]
    \centering
    \includegraphics[width=0.98\textwidth]{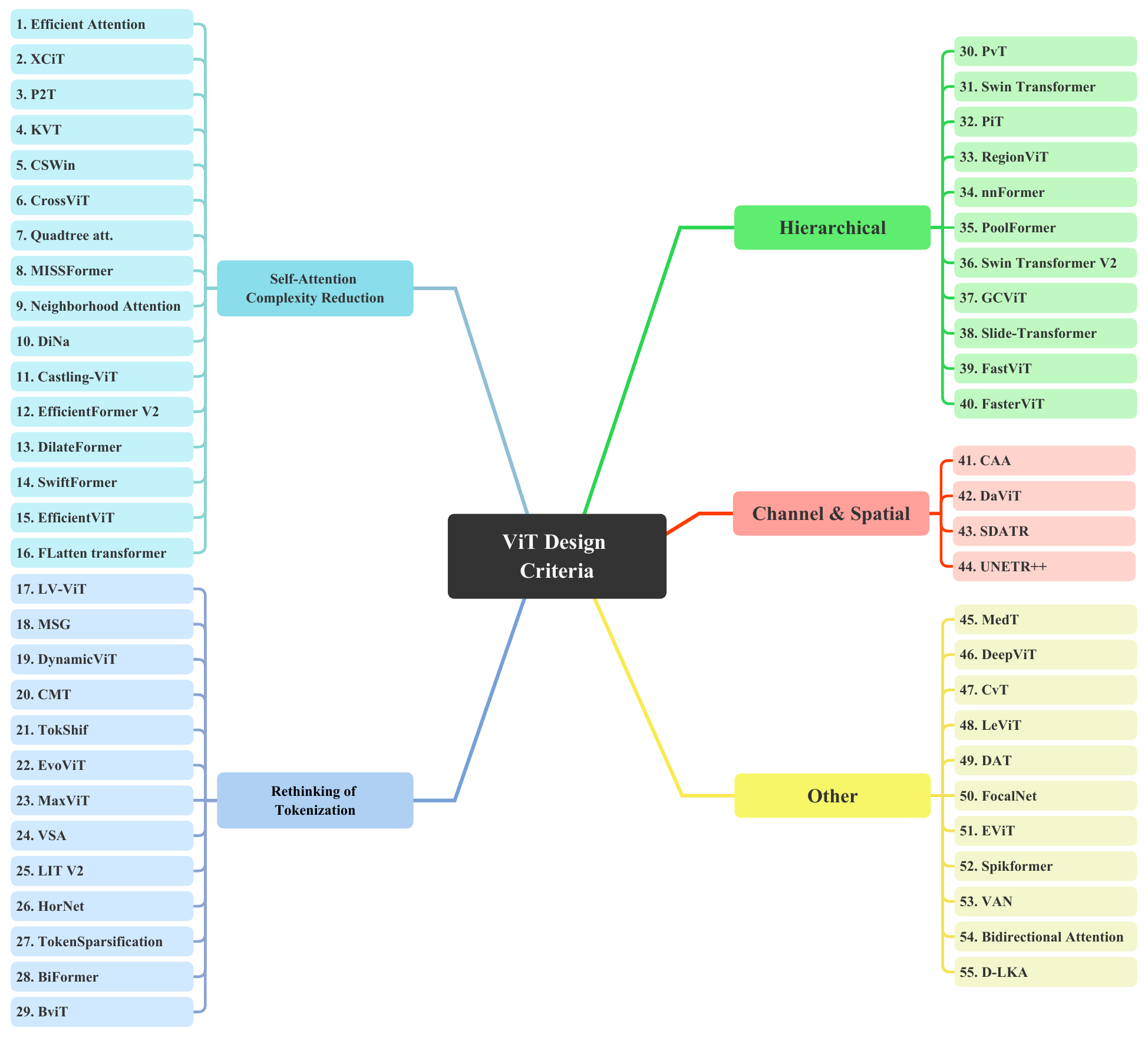}
	\caption{The suggested taxonomy for attention mechanisms used within ViTs consists of four distinct groups: I) Computation Reduction, II) Hierarchical, III) Channel \& Spatial, IV) Other. To maintain conciseness, we assign ascending prefix numbers to each category in the paper's name and cite each study accordingly as follows: \protect\input{extra/taxonomy-references}}
    \label{fig-taxonomy-design}
\end{figure*}

The nomenclature used here is in contrast to the unified attention model~\cite{niu2021review}, where queries and keys are always multiplied first to receive the
attention weights. But it is also stated in~\cite{niu2021review} that query, key, value are arbitrary representations of the input features, and therefore the names
can be interchanged to fit the unified model.

\subsection{XCiT - Cross-Covariance Image Transformer}

Ali et al.~\cite{ali2021xcit} propose the XCiT, a cross-covariance based ViT.

A major problem with self attention is the quadratic complexity relative to the number of input tokens.
XCiT alleviates the problem by introducing cross-covariance attention:

\begin{align}
	\mathdash{XC-Attention}(\mathbf{Q},\mathbf{K},\mathbf{V}) & = \mathbf{V} A_{XC}(\mathbf{K},\mathbf{Q}),  \nonumber               \\
	A_{XC}(\mathbf{K},\mathbf{Q})                             & = \operatorname{softmax}(\hat{\mathbf{K}}^{T}\hat{\mathbf{Q}}/\tau).
\end{align}
\begin{figure}
	\includegraphics[width=\linewidth]{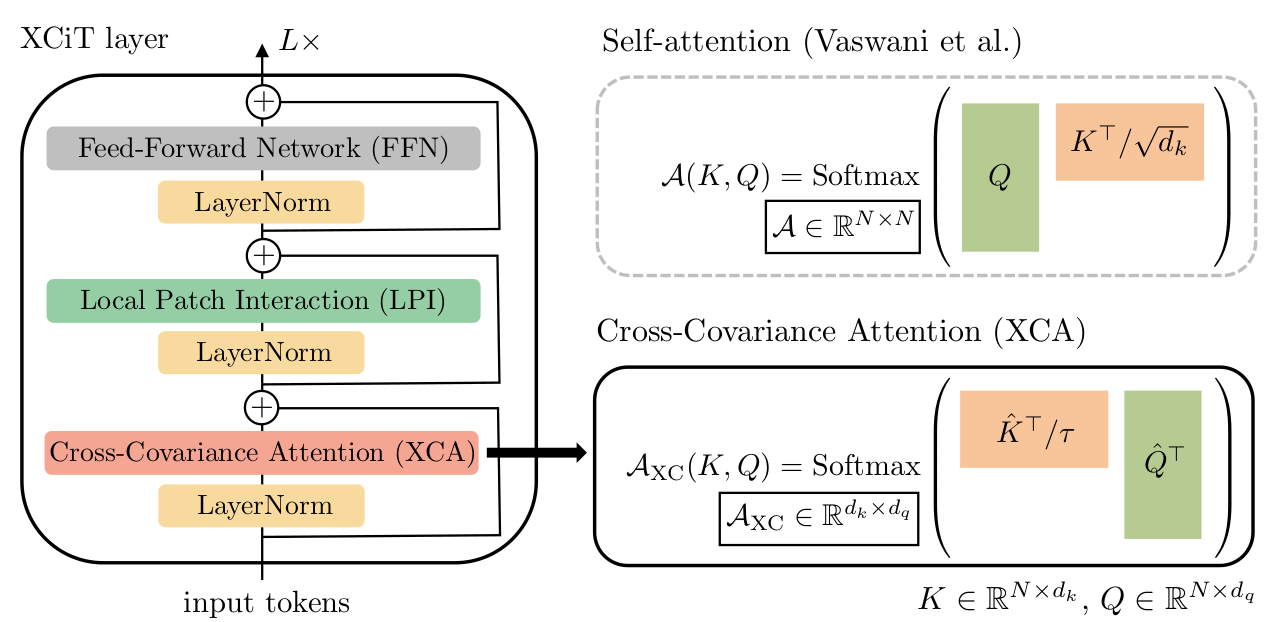}
	\caption{Regular self attention (top right) and cross-covariance attention (bottom right). From~\cite{ali2021xcit}.}\label{fig-cross-covariance}
\end{figure}
A comparison is shown in \Cref{fig-cross-covariance}. The keys and queries are transposed, therefore the attention weights
are based on the cross-covariance matrix. The temperature parameter $\tau$ is introduced to counteract the scaling with the $l_{2}$-norm that is applied
to the queries and keys before calculating the attention weights. This increases stability during training but removes a degree of freedom, thus limiting the representational capability of the network.
The complexity of cross-covariance attention and self attention is compared in \Cref{tab-attention-complexity}. $N$ refers to the number of tokens,
$h$ is the number of heads and $d$ is the feature dimension.
Because the keys and queries are transposed, cross-covariance attention is a channel attention mechanism.

The XCiT excels at handling larger images ($>$1000 pixels per dimension), which regular ViT does not because of the large number of patch tokens resulting from
the image size.

\subsection{CrossViT - Cross Attention Multi-Scale Vision Transformer}

Based on the success of the ViT, Chen et al.~\cite{chen2021crossvit} introduce the cross-attention multi-scale Vision Transformer (CrossViT).
It improves the accuracy and - more importantly - the performance of the ViT. This method employs both spatial attention as used in the ViT and branch attention. In this case, a branch refers to image patches at different scales. 

CrossViT utilizes two different patch sizes for its images, one large patch main branch (\textit{L-Branch}) and a small complementary branch (\textit{S-Branch}). The large branch computes larger patch sizes, but has more encoders and wider embedding dimensions
than the small branch. In both branches, patches are linearly projected and a classification token (\textit{cls} token) is added, like in the ViT. Transformer encoders process each branch separately. Next,
the resulting tokens are fused with cross attention. Afterwards, The two cls tokens are processed by one MLP each. The result is added for the classification.
Chen et al. tested several fusion techniques:

\begin{itemize}
	\item All attention fusion - self-attention over all tokens ($O(N^{2})$)
	\item Class token fusion - only the class tokens are fused ($O(1)$)
	\item Pairwise fusion - pairs of tokens are fused based on the spatial location ($O(N)$)
	\item Cross attention - cls token of one branch fused with class tokens of the other ($O(N)$)
\end{itemize}
Since all attention requires quadratic computation time relative to the number of tokens, a more efficient method is presented.
This method is called cross-attention fusion. The cls token of one branch is compared - via the attention mechanism - to the patch tokens of the other branch and vice versa.

The cls token is used as the query token for attention:

\begin{align}
	\mathbf{x}'^{l} = [f^{l}(\mathbf{x}^{l}_{cls}) || \mathbf{x}^{s}_{patch}],
	\label{eq-cross-attention-xl}
\end{align}
where $f^{l}(\cdot)$ is a projection to align the dimensions of small and large patches. The cross attention can then be expressed as:

\begin{align}
	\mathbf{Q} = \mathbf{x}'^{l}_{cls} \mathbf{W}_{Q}, \mathbf{K} = \mathbf{x}'^{l}\mathbf{W}_{K}, \mathbf{V} = \mathbf{x}'^{l} \mathbf{W}_{V}, \\
	\mathbf{A} = \operatorname{softmax}(\mathbf{QK}^{T}/\sqrt{C/h}), CA(\mathbf{x}'^{l} = \mathbf{AV}).
	\label{eq-cross-attention}
\end{align}
$\mathbf{W}_{Q}, \mathbf{W}_{K}, \mathbf{W}_{V} \in \mathbb{R}^{Cx(C/h)}$ are learnable parameters, $C$ is the embedding dimension and $h$ is the number of heads.

The output of the whole cross-attention module is defined as follows:

\begin{align}
	\mathbf{y}^{l}_{cls} & = f^{l}(\mathbf{x}^{l}_{cls}) + \operatorname{MCA}(\operatorname{LN}([f^{l}(\mathbf{x}^{l}_{cls}) || \mathbf{x}^{s}_{patch}])), \\
	\mathbf{z}^{l}       & = [g^{l}(\mathbf{y}^{l}_{cls}) || \mathbf{x}^{l}_{patch}],
	\label{eq-cross-attention-module}
\end{align}
with $\operatorname{MCA}$ being multi-head cross attention and $\operatorname{LN}$ being layer normalization.

The main advantage of CrossViT is a more efficient model because the number of transformer encoders is small for the small branch patches.
Unlike ViT, CrossViT also performs well on tasks with small datasets for training.

\subsection{EdgeNext}

EdgeNext is an architecture proposed by Maaz et al.~\cite{maaz2022edgenext} for edge devices. It is specifically optimized to reduce the number of Multiplication-Addition (MAdd) operations
required.
It uses an attention mechanism similar to the XCiT, called split depth-wise transpose attention (SDTA).

The input is split into $s$ subsets of the same size.
A $3 \times 3$ depth-wise convolution processes each subset. The stage number $t$, where $t \in {1, 2, 3, 4}$ determines the number of subsets dynamically.
In order to have linear complexity in the number of tokens, cross-covariance attention, also called transpose attention, is applied afterward. 

This is a form of channel attention since the attention is now applied to the channel dimension of the input due to the transpose operation.

\subsection{MISSFormer}

The MISSFormer by Huang et al.~\cite{huang2022missformer}, introduces efficient self-attention (ESA) and the enhanced transformer context bridge.


The MISSFormer applies a hierarchical structure with efficient self-attention blocks along with a multiscale fusion technique referred to as the enhanced transformer context bridge.
It also employs a U-Net-like structure of an encoder and a decoder, both working with transformer blocks only.

Efficient self-attention is a spatial attention mechanism that makes use of spatial reduction, represented by the spatial reduction ratio $R$.
The number of tokens $N$ is reduced by $R$ while the channel dimension is expanded by $R$. The complexity
of ESA is reduced to $O(\frac{N^{2}}{R})$ whereas unmodified self attention is $O(N^{2})$.

The ESA can be written as:

\begin{align}
	\operatorname{Attention}(\mathbf{Q}, \mathbf{K}, \mathbf{V}) = \operatorname{softmax}(\frac{\mathbf{QK^{T}}}{\sqrt{d_{head}}}\mathbf{V}), \\
	\mathbf{K} = \operatorname{Reshape}(\frac{N}{R}, C \cdot R) W (C \cdot R, C).
\end{align}
$K$ and $V$ are reshaped to $\frac{N}{R} \times (C \cdot R)$, reducing the spatial dimension by the reduction ratio $R$. A linear projection $W$ is employed to regain the channel depth C.

The enhanced transformer context bridge fuses information of different hierarchical levels by first concatenating the feature tokens from all levels,
calculating ESA on the merged tokens, then splitting the tokens up again.
The split token sequence is transformed to image patches and a feed-forward network called Enhanced Mix-FFN is applied to each hierarchical level. These are again tokenized, concatenated,
and lastly fed back to the decoder of the MISSFormer.

\subsection{SwiftFormer}

SwiftFormer ~\cite{shaker2023swiftformer} introduces an innovative efficient additive attention mechanism, replacing quadratic matrix multiplication operations with linear element-wise multiplications. This design affirms the substitutability of the key-value interaction with a linear layer without compromising accuracy.

Unlike traditional additive attention mechanisms in NLP, which capture global context through pairwise interactions between tokens via element-wise multiplications instead of dot-product operations, it is demonstrated that removing key-value interactions while focusing solely on effectively encoding query-key interactions with a linear projection layer is sufficient. Termed \textit{"efficient additive attention"}, this approach exhibits faster inference speeds and yields more robust contextual representations, as evidenced by notable performance improvements in main and downstream CV tasks.

To delve into specifics, the transformation of the input embedding matrix \(x\) into query (\(Q\)) and key (\(K\)) employs two matrices \(W_q\) and \(W_k\), where \(Q, K \in \mathbb{R}^{n \times d}\), \(W_q, W_k \in \mathbb{R}^{d \times d}\), \(n\) is the token length, and \(d\) is the dimensionality of the embedding vector. The subsequent multiplication of the query matrix \(Q\) by the learnable parameter vector \(w_a \in \mathbb{R}^d\) generates attention weights for the query, producing the global attention query vector \(\alpha \in \mathbb{R}^n\) as:

\begin{align}
    \alpha = Q \cdot w_a/\sqrt{d} 
\end{align}

The query matrix is then pooled based on the learned attention weights, resulting in a single global query vector \(q \in \mathbb{R}^d\) given by:

\begin{align}
    q = \sum_{i=1}^{n} \alpha_i \ast Q_i
\end{align}

Subsequently, interactions between the global query vector \(q \in \mathbb{R}^d\) and the key matrix \(K \in \mathbb{R}^{n \times d}\) are encoded using the element-wise product, forming the global context \(\mathbb{R}^{n \times d}\). This matrix, akin to the attention matrix in Multi-Head Self Attention (MHSA), captures information from every token and exhibits flexibility in learning correlations within the input sequence.

Drawing inspiration from the transformer architecture, a linear transformation layer is employed for query-key interactions to learn the hidden representation of tokens. The output of the efficient additive attention, denoted as \(\hat{x}\), is described by:

\begin{align}
    \hat{x} = \hat{Q} + T(K \ast q)
\end{align}

where \(\hat{Q}\) represents the normalized query matrix, and \(T\) signifies the linear transformation.

\section{Hierarchical Transformers} \label{hierarchical}

In the next section, hierarchical transformer architectures are shown.

\subsection{Swin - Hierarchical Vision Transformer Using Shifted Windows}

Liu et al.~\cite{liu2021swin} introduce a transformer with two new concepts: a hierarchical feature map scheme and an attention mechanism with shifted windows.

In the first stage, the input image patches are of size $\frac{H}{4} \times \frac{W}{4} \times 48$, with $H, W$ being the input height and width, respectively.
After each stage, $2\times2$ patches are merged into one patch to gain a hierarchical representation.

The two Swin Transformer blocks in each stage use windowed multi-head self-attention (W-MSA) and shifted window MSA.

Shifted window self-attention is a spatial attention mechanism. It operates on local windows for efficiency - the complexity is still quadratic with regard to
the number of patches, but the number of patches is small due to attention being restricted to local windows. To model connections across windows, the Swin approach alternates between two shifted configurations. The second configuration is displaced by
half the window size. Each Swin Transformer block is followed by a shifted Swin Transformer block. 

To improve the computation of the shifted window, which is composed of many non-quadratic parts, a cyclic shift is applied together with a masked MSA. This keeps the number of batched windows
the same as in the standard window configuration.
The Swin-T performs similarly to state-of-the-art CNNs like ResNet-152~\cite{resnet}.

\subsection{RegionViT - Region Vision Transformer}

Chen et al.~\cite{chen2021regionvit} introduce regional-to-local attention in their paper RegionViT. The advantage lies in the reduced complexity
by $O(N/M^{2})$, where $N$ is the number of tokens and $M$ is the window size.

Regional-to-local attention is a combination of two spatial attention mechanisms -
Regional self-attention (RSA) uses regional tokens to exchange information between regions and local self-attention (LSA) is the same as self-attention in the ViT~\cite{dosovitskiy2020vit}.
To reduce the number of parameters, RSA and LSA share their weights.
Essentially, RSA and LSA are tokenized the same way, but RSA tokens have a larger patch size, hence each RSA token covers the region of $7^{2}$ LSA tokens. Both are tokenized
by convolution.

\subsection{GCViT - Global Context Vision Transformers}

Hatamizadeh et al.~\cite{hatamizadeh2023global} present the Global Context Vision Transformer, which employs a twofold attention to generate local and global context, respectively.

Local attention is computed on local window patches, whereas the global queries are generated via the global query generator and attention is calculated
between global queries and local key and value tokens.

The global query generator works as follows:

\begin{align}
	\mathbf{x}^{i} = \mathdash{f-MBConv}(\mathbf{x}^{i-1}), \nonumber \\
	\mathbf{x}^{i} = \operatorname{MaxPool}(\mathbf{x}^{i}).
\end{align}
$i \in \{1, 2, 3, 4\}$ refers to the stage.
$\mathdash{f-MBConv}$ refers to modified fused inverted residual blocks:

\begin{align}
 \mathbf{x} = \operatorname{Conv}_{1 \times 1}(\operatorname{SE}({\operatorname{GELU}({\mathbf{\mathdash{DW-Conv}_{3 \times 3}(\mathbf{x})}})})) + \mathbf{x},
\end{align}
where $\mathdash{DW-Conv}$ refers to depth-wise convolution and $\operatorname{SE}$ is a squeeze-and-excitation block~\cite{SEnet}. $\operatorname{GELU}$ denotes the gaussian error linear unit~\cite{hendrycks2016gelu}.
$\mathbf{x}$ is the input tensor.

The global tokens are a way to generate global context and the global attention enables the local tokens to ``see''
the global context by multiplying global query tokens and local key tokens to compute the attention weights.

\subsection{nnFormer}
The authors of nnFormer \cite{zhou2023nnformer} present a 3D transformer network, that utilizes local and global attention as well as a combination of interleaved convolution and self-attention. \\
The network architecture consists of three parts: the encoder, the bottleneck, and the decoder. In the encoder convolutional layers are used as an embedding layer to precisely encode spatial information and capture low-level features. Local window self-attention is used to capture long-range dependencies in an efficient manner for high-resolution inputs. In contrast to Swin Transformer \cite{liu2021swin}, a volumetric instead of a two-dimensional input is used. After each local attention layer convolutional down-sampling is used to reduce the size of the feature maps. In the bottleneck, the feature size is small enough to use global self-attention without increasing the computational costs by a large amount. With global attention high level, long-range dependencies are captured.\\
In the bottleneck, three global attention layers are used. In the decoder, the features are up-sampled to restore the full feature size. Similar to the encoder, local window attention is used again. 
In the skip connections skip attention is used. The skip attention combines information from the encoder side, represented by features that are projected to the keys and values via a linear layer. The information is fused with information from the decoder, represented by the queries. Both are combined either via local or global attention. The final output is produced by an expanding layer, that restores the original input resolution.


The downside of the approach is the high number of FLOPs (213.4 G). Also, the limitations for local window attention are similar to those for the Swin Transformer.

\subsection{Fast Vision Transformers with Hierarchical Attention}
Hatamizadeh et al. propose FasterViT ~\cite{hatamizadeh2023fastervit},  a novel hybrid CNN-ViT neural network, that focuses on optimizing image throughput for CV applications. Combining the advantages of fast local representation learning from CNNs and global modeling properties inherent in ViTs, FasterViT introduces a Hierarchical Attention (HAT) approach. HAT effectively decomposes global self-attention with quadratic complexity into a multi-level attention system, significantly reducing computational costs. 

The model utilizes efficient window-based self-attention, where each window has dedicated carrier tokens contributing to both local and global representation learning. At a higher level, global self-attentions facilitate efficient cross-window communication at reduced costs.

FasterViT comprises four stages, involving a reduction in input image resolution through a strided convolutional layer while doubling the number of feature maps. The design incorporates residual convolutional blocks~\cite{resnet,SEnet} in early high-resolution stages (Stage 1, 2) and transformer blocks in later stages (Stage 3, 4). This strategy enables the rapid generation of high-level tokens, further processed using transformer-based blocks. Each transformer block follows an interleaved pattern of local and newly proposed Hierarchical Attention blocks, effectively capturing short and long-range spatial dependencies and efficiently modeling cross-window interactions. The proposed Hierarchical Attention efficiently learns \textit{carrier tokens} as summaries of each local window, facilitating efficient cross-interaction between regions. Despite the computational complexity of Hierarchical Attention growing nearly linearly with input image resolution, it proves to be an efficient and effective approach for capturing long-range information with large input features.

In this study, a novel formulation of windowed attention is proposed, building upon local windows introduced in the Swin Transformer~\cite{cao2022swin,liu2021swinv2}. The introduction of \textit{carrier tokens} (CTs) serves to play the summarizing role for entire local windows. The initial attention block applies to CTs to \textit{summarize} and \textit{propagate} global information. Subsequently, local window tokens and CTs are \textit{concatenated}, ensuring each local window exclusively accesses its set of CTs. By employing self-attention on concatenated tokens, local and global information exchange is facilitated at a reduced cost. An alternation between sub-global (CTs) and local (windowed) self-attention formulates the concept of hierarchical attention. Conceptually, CTs can be further grouped into windows, creating a higher order of carrier tokens.

\section{Channel and Spatial Transformer Architectures}\label{ch-spatial}

\subsection{DaViT - Dual Attention Vision Transformer}

The Dual Attention Vision Transformer (DaViT) by Ding et al.~\cite{ding2022davit} combines spatial and channel attention.

\begin{figure}
	\includegraphics[width=\columnwidth]{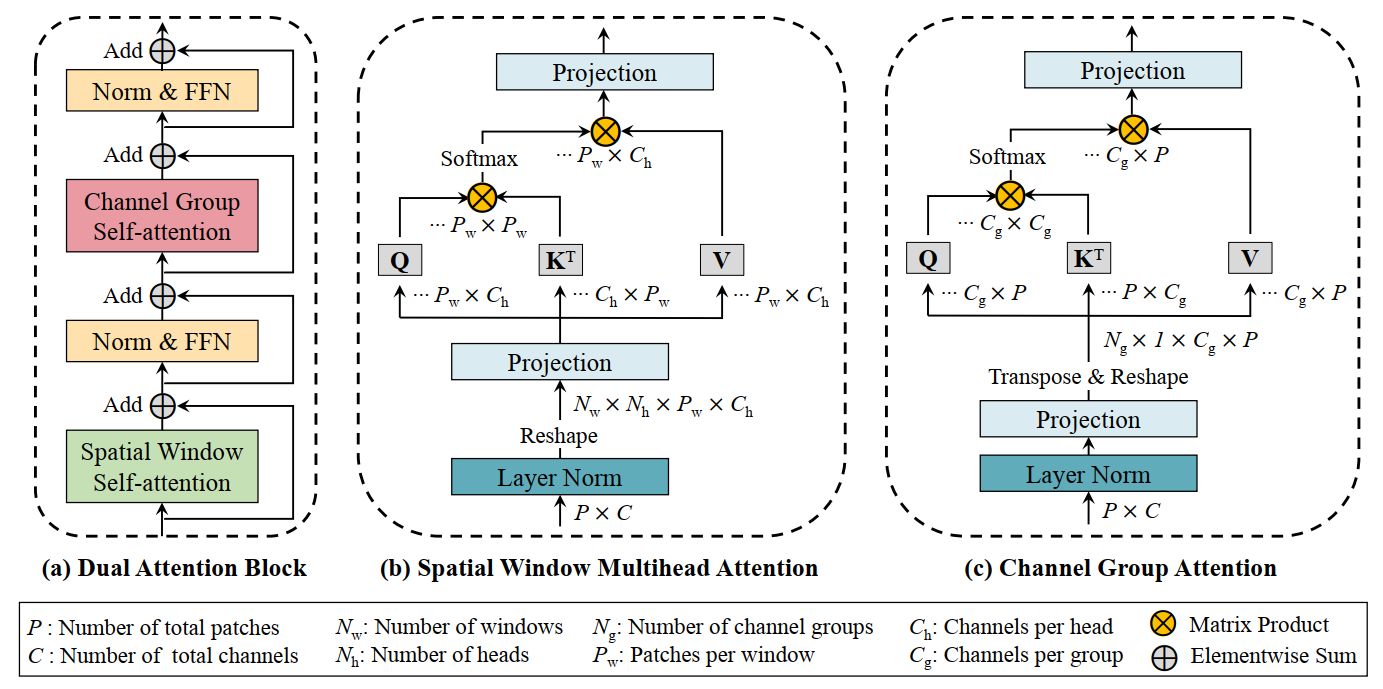}
	\caption{The DaViT dual attention block with spatial and channel attention. From~\cite{ding2022davit}.}\label{fig-dual-attention-transformer}
\end{figure}

The dual attention transformer tackles the issue of global context versus complexity. Previous approaches either reduce the complexity but lose global contextual information
or are affected by the quadratic complexity of the self-attention mechanism.

The combination of spatial and channel attention counteracts the aforementioned problem by combining spatial window attention and channel group attention.
The latter allows the model to still capture global relationships while the former stays linear in complexity relative to the spatial dimension.

Spatial window attention can be expressed as follows:

\begin{align}
	A_{window}(\mathbf{Q},\mathbf{K},\mathbf{V}) = \{A(\mathbf{Q}_{i},\mathbf{K}_{i},\mathbf{V}_{i})\}_{i=0}^{N_{w}},
\end{align}
where $\mathbf{Q}_{i},\mathbf{K}_{i},\mathbf{V}_{i} \in \mathbb{R}^{P_{w} \times C_{h}}$ denote local window queries, keys, and values, respectively.
$N_{w}$ refers to the number of different windows. This window attention cannot model global contextual information, which is solved by channel group attention.

The feature tokens resulting from window attention are transposed. The transposed tokens are grouped for reduced complexity and the channel attention is calculated:

\begin{align}
	A_{channel}(\mathbf{Q},\mathbf{K},\mathbf{V}) = \{A_{group}(\mathbf{Q}_{i},\mathbf{K}_{i},\mathbf{V}_{i})^{T} \}_{i=0}^{N_{g}}, \nonumber \\
	A_{group}(\mathbf{Q}_{i},\mathbf{K}_{i},\mathbf{V}_{i}) = \operatorname{softmax} \left(\frac{\mathbf{Q}_{i}^{T} \mathbf{K}_{i}}{\sqrt{C_{g}}} \right)  \mathbf{V}_{i}^{T}.
\end{align}
$N_{g}$ refers to the number of groups and $C_{g}$ to the number of channels per group.
$\mathbf{Q}_{i},\mathbf{K}_{i},\mathbf{V}_{i} \in \mathbb{R}^{P \times C_{g}}$ are grouped channel-wise queries, keys, and values.

The dual attention block implemented by the DaViT is shown in \Cref{fig-dual-attention-transformer}. A dual attention block employs spatial window self-attention followed by normalization and a fully connected layer, which is fed
to the channel group self-attention, again followed by a normalization layer. Each sublayer has a residual connection around it.

\subsection{Spatial Spectral Transformer}

\begin{figure}
	\includegraphics[width=\columnwidth]{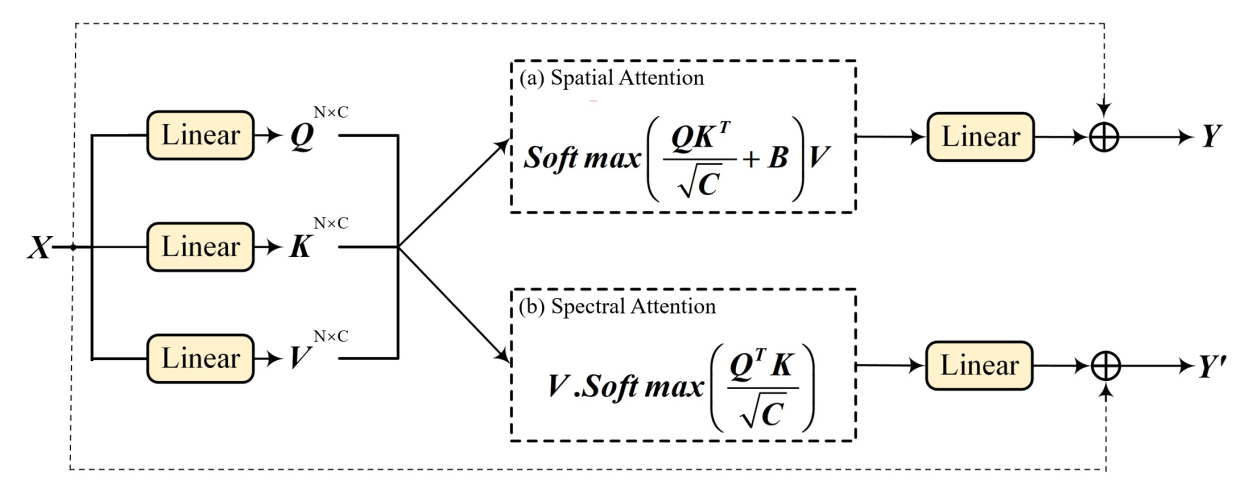}
	\caption{Spatial attention in (a) and spectral attention in (b). From~\cite{sun2022fusing}.}\label{fig-spatial-spectral-attention}
\end{figure}

Sun et al.~\cite{sun2022fusing} introduce another dual attention transformer for remote sensing, a transformer using spatial attention together with channel attention on spectral images.
The purpose of this architecture is to classify hyperspectral images (HSI). The two attention mechanisms used are shown in~\Cref{fig-spatial-spectral-attention}.

Long-range spatial context is encoded by spatial attention, whereas channel attention is used to gain information from the spectral depth.
Three methods of fusing channel and spatial attention are explored:
\textit{Additive}, \textit{concatenated} and \textit{multiplicative} fusion. 
Concatenated feature fusion performs the best out of the three approaches, according to Sun et al.~\cite{sun2022fusing}.

Their proposed transformer network uses hierarchical shifted-window attention like the Swin-T and spectral channel attention in each transformer block.

\subsection{SCViT - Spatial Channel Vision Transformer}

Lv et al.~\cite{lv2022scvit} propose the SCViT, a spatial-channel Vision Transformer for remote sensing.
It employs regular transformer blocks with multi-head self-attention as its backbone.
The token generation is performed by the progressive token aggregation module (PA module)~\cite{yuan2021tokens}.
The lightweight channel attention (LCA) module is used for classification. An LCA block reweighs the channels of the classification token $\mathbf{t}_{cls}$:

\begin{align}
	\mathbf{y} = \operatorname{softmax}(\operatorname{FC}(\operatorname{LCA}(\mathbf{t}_{cls}))).
\end{align}
The LCA module reweighs channels by using a 1D convolution. The reweighed cls token is then run through a fully connected layer followed by a softmax for classification.
Channel attention is applied to the classification token to leverage channel information which is important for the classification task.

\subsection{CAA - Channelized Axial Attention}

Huang et al.~\cite{huang2022channelized} propose channelized axial attention, a dual attention mechanism that seamlessly combines spatial and channel attention in one operation.
According to Huang et al., the problem with parallel and sequential dual attention is that spatial and channel attention may have conflicting features that may block the useful results from one operation.
Huang et al. therefore propose to calculate channel attention inside axial attention, postulating that channel attention does not require the whole feature map to compute useful outputs.


In axial attention, the spatial domain is split into rows $A_{row}$ and columns $A_{col}$ and attention is performed separately on each. 

To simplify the dual attention, the final attention is shortened:

\begin{align}
	\boldsymbol{\alpha} & = A_{col}(\mathbf{x}_{i,j},\mathbf{x}_{m,j})g(\mathbf{x}_{m,n}),                  \\
	\boldsymbol{\beta}  & = A_{row}(\mathbf{x}_{i,j},\mathbf{x}_{i,n}) \sum_{\forall m}\boldsymbol{\alpha}, \\
	\mathbf{y}_{i,j}    & = \sum_{\forall n} \boldsymbol{\beta}.
\end{align}
Channel attention is now seamlessly integrated into the module using spatially varying channel attention:

\begin{align}
	C_{col}(\boldsymbol{\alpha}) & = \operatorname{Sigmoid}\left(\operatorname{ReLU}(\frac{\sum_{\forall m,j}(\boldsymbol{\alpha})}{H \times W}\omega_{c1})\omega_{c2} \right)\boldsymbol{\alpha}, \\
	C_{row}(\boldsymbol{\beta})  & = \operatorname{Sigmoid}\left(\operatorname{ReLU}(\frac{\sum_{\forall i,n}(\boldsymbol{\beta})}{H \times W}\omega_{r1})\omega_{r2} \right)\boldsymbol{\beta},
\end{align}
where $\omega_{c1}, \omega_{c2}, \omega_{r1}, \omega_{r2}$ are learnable weights. The output of the channelized attention module is:

\begin{align}
	\mathbf{y}_{i,j} = \sum_{\forall n}C_{row} \left(A_{row}(\mathbf{x}_{i,j},\mathbf{x}_{i,n})(\sum_{\forall m}C_{col}(\boldsymbol{\alpha})) \right).
\end{align}
Channel attention is computed for each row separately.

\subsection{Semantic-Enhanced Dual Attention}

Semantic-enhanced dual attention transformer (SDATR) is a network proposed by Ma et al.~\cite{ma2022knowing}. It is a transformer architecture designed for image captioning tasks, i.e. assigning
a description to an image. 


The spatial attention is standard multi-head self-attention. Channel attention first performs channel reduction with a 1x1 convolution, then applies global average pooling to aggregate spatial information in each channel.
Afterward, a gating mechanism is utilized to obtain the attention weights of each channel. These weights are then applied to the reduced visual feature.



The architecture utilizes faster R-CNN~\cite{ren2015faster} to generate grid feature maps. These are input to the transformer encoder, employing dual attention modules and feed-forward networks with residual connections.
The decoder also processes text information, therefore the features of the encoder and the embedding of the description text are cross-attended. The output is a description fitting the image.
This method adds the ability to learn descriptive characteristics of the input image to the existing capabilities of the ViT - capturing long-range context in images.

\subsection{UNETR++}
Shaker et al. \cite{shaker2022unetr++} present and efficient network for accurate 3D medical image segmentation. It combines channel and spatial attention in a paired attention block.\\
First, the input volume $\mathbf{x}\in \mathcal{R}^{H\times W \times D}$ is divided into non-overlapping patches. The network consists of multiple Efficient Paired Attention (EPA), that are placed in a U-shaped manner. After each encoder stage, the resolution is halved. In the decoder, it is doubled.\\
The EPA block combines effective spatial attention with channel attention to learn rich features in both the spatial and channel dimensions. The weights of the query $Q$ and key $K$ linear layers are shared between the two attention modules. By sharing weights complementary features between the two types of attention are learned. This results in better feature representations and fewer parameters. A unique value layer $V$ is learned for each attention method. \\
In spatial attention, the token dimension $n$ is reduced to a projection dimension $p$ with $p << n$ for the keys $K_{shared}$ and values $V_{spatial}$. The complexity is reduced from $O\left(n^2\right)$ to $O\left(np\right)$. Self-attention is performed with the projected key and value and the shared query matrices:
\begin{equation}
    \hat{\mathbf{X}}_p = Softmax\left(\frac{\mathbf{Q}_{shared}\mathbf{K}^{T}_{proj}}{\sqrt{d}}\right) \cdot \Tilde{\mathbf{V}}_{spatial}.
\end{equation}
Here, $\Tilde{\mathbf{V}}_{spatial}$ are the projected spatial values and $d$ is the length of each vector.\\
In channel attention, the dependencies between different channels of the feature maps are captured.  Again, the shared query $Q_{shared}$ and keys $K_{shared}$ are used, while a unique value $V_{channel}$ is received from a linear layer. The channel attention is shown in the following equation:
\begin{equation}
    \hat{\mathbf{X}}_c = \mathbf{V}_{channel} \cdot Softmax \left(\frac{\mathbf{Q}_{shared}^{T}\mathbf{K}_{shared}}{\sqrt{d}}\right).
\end{equation}
In contrast to self-attention, the order of multiplications of the keys and queries and the values and the similarity matrix are swapped. This results in reduced computations. The features of the two branches are fused. A richer feature representation is generated by additional convolution blocks.\\
UNETR++ effectively combines spatial and channel dimensions and achieves excellent results in multiple datasets.

\section{Transformers Rethinking Tokenization}\label{rethinking}

In the next section, state-of-the-art transformer architectures are presented that expand tokenization in different ways.

\subsection{DynamicViT - Dynamic Vision Transformer}

Rao et al.~\cite{rao2021dynamicvit} introduce a transformer architecture that applies dynamic token sparsification. It specifically aims at reducing model complexity and speeding up inference times by learning
which tokens are more relevant to the network's prediction. This is similar to CNN models that remove redundant filters. 

The token sparsification happens hierarchically, i.e. after every transformer block, tokens are dropped based on a binary decision mask $\mathbf{\hat{D}} \in \{0,1\}^{N}$, with $N$ being the number of
tokens. First, all values in $\mathbf{\hat{D}}$ are set to 1. Then, the current decision is updated by sampling from a distribution $\mathbf{\pi}$:

\begin{align}
	\mathbf{\pi} = \operatorname{softmax}(\operatorname{MLP}(\mathbf{z})) \in \mathbb{R}^{N \times 2},
\end{align}
where $\mathbf{z}$ is a combination of local and global features learned from two separate MLPs applied to the input feature $\mathbf{x}$, and in case of the global feature, an aggregation with the decision mask $\mathbf{\hat{D}}$.
The current decision is then generated as follows:

\begin{align}
	\mathbf{\hat{D}} \leftarrow \mathbf{\hat{D}} \odot \mathbf{D},
\end{align}
where $\odot$ is the Hadamard product (elementwise multiplication).

DynamicViT greatly reduces the number of tokens and increases the throughput while only suffering a minor reduction in accuracy.

\subsection{MSG Transformer - Exchanging Local Spatial Information by Manipulating Messenger Tokens}

Fang et al.~\cite{fang2022msg} introduce the MSG transformer. It utilizes message tokens to send information between local windows.

In the MSG transformer, a hierarchical structure is used along with window attention. The resulting patch tokens are then expanded by a
messenger token (MSG token), which is used to exchange information in a shuffle region. 

The novel part here is the idea of messenger tokens and the shuffle operation, which exchanges the messenger tokens between local windows.
The MSG transformer reduces the computational complexity of the transformer network by limiting the spatial attention to local windows. Instead of shifting windows,
information exchange is done via messenger tokens.


\subsection{All Tokens Matter}

Jiang et al.~\cite{jiang2021all} present a novel token labeling scheme where not only the cls token, but all patch tokens carry classification information as well.

The labels assigned to each patch token are stored in a dense score map. The output patch token and related label are used to calculate the cross-entropy loss, which
is applied as an auxiliary loss during the training phase.
The token labeling objective is:
\begin{align}
	L_{tl} = \frac{1}{N} \sum_{i=1}^{N} H(\mathbf{X}^{i}, y^{i}).
\end{align}
The total loss then becomes:

\begin{align}
	L_{total} & = H(\mathbf{X}^{cls}, y^{cls}) + \beta \cdot L_{tl},                                             \\
	& = H(\mathbf{X}^{cls}, y^{cls}) + \beta \cdot \frac{1}{N} \sum_{i=1}^{N} H(\mathbf{X}^{i}, y^{i}).
\end{align}
$H(\cdot)$ refers to the softmax cross-entropy loss and $y^{cls}$ to the class label. $\beta$ is set to 0.5.
These token labels provide additional location-specific information for each patch. The operations required to match the dense score map to the target image are negligible compared to the attention mechanism in the transformer blocks.
As the score map is dense, it fits well to downstream tasks like semantic segmentation.

\subsection{TokShift - Token Shift Transformer}

\begin{figure}
	\begin{subfigure}{0.24\textwidth}
		\centering
		\includegraphics[width=0.8\linewidth]{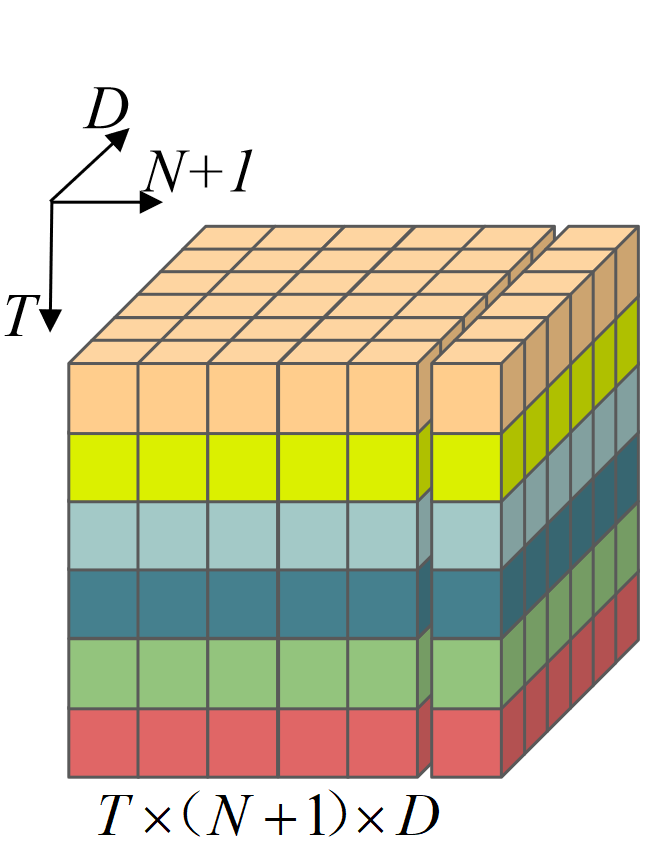}
	\end{subfigure}
	\begin{subfigure}{0.24\textwidth}
		\centering
		\includegraphics[width=0.8\linewidth]{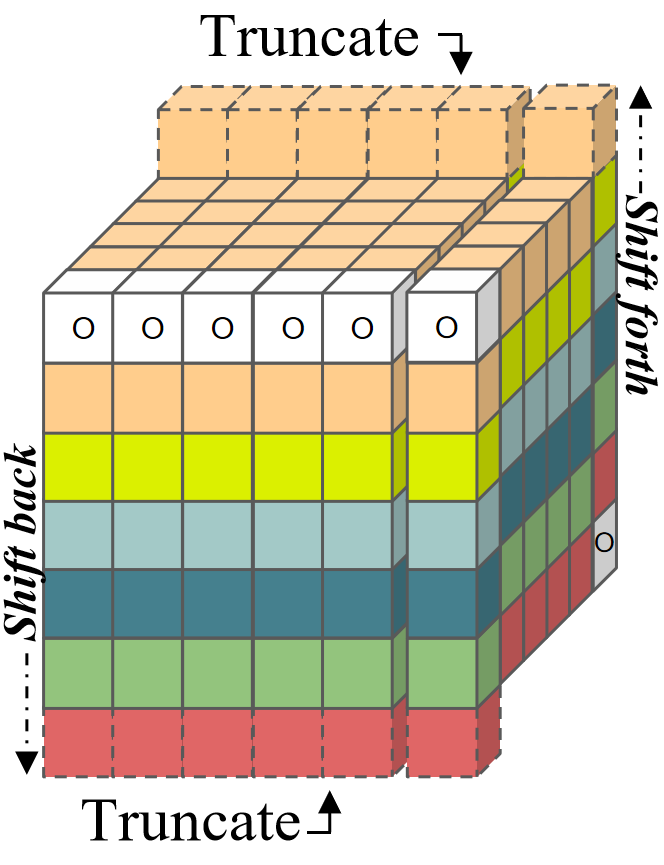}
	\end{subfigure}
	\begin{subfigure}{0.24\textwidth}
		\centering
		\includegraphics[width=0.8\linewidth]{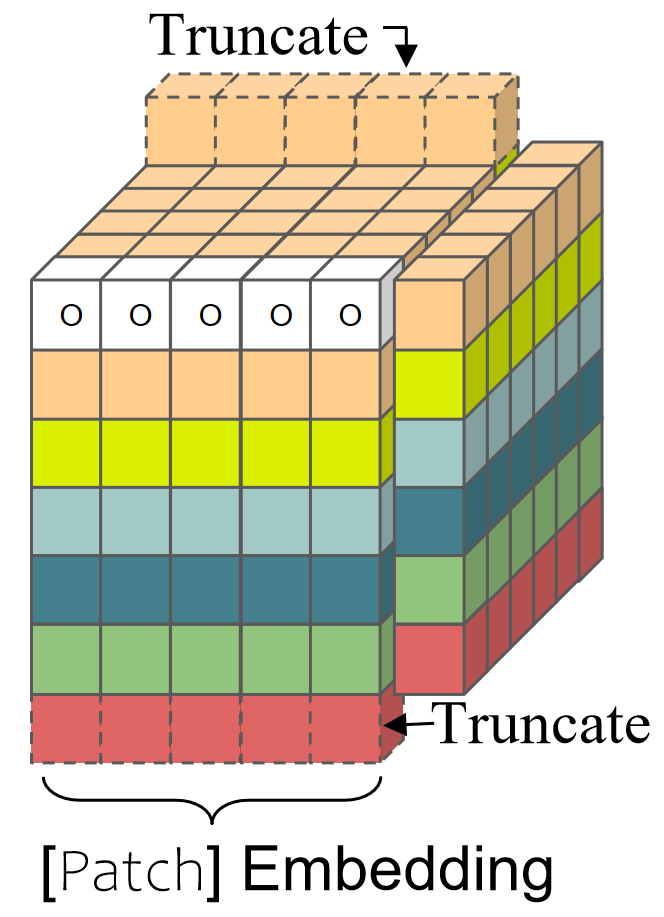}
	\end{subfigure}
	\begin{subfigure}{0.24\textwidth}
		\centering
		\includegraphics[width=0.8\linewidth]{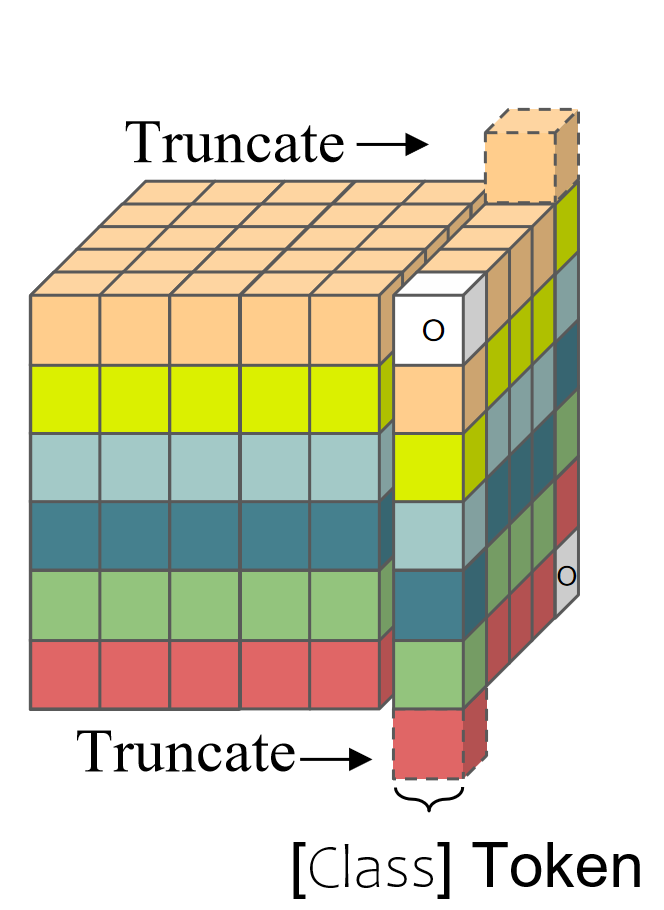}
	\end{subfigure}
	\caption{Token shift operations. Either do not shift (a), shift both cls and patch token (b), shift only the patch token (c), or only the cls token (d). From~\cite{zhang2021token}.}\label{fig-token-shift-transformer}
\end{figure}

The token shift transformer is introduced by Zhang et al.~\cite{zhang2021token}. It operates on video data, hence a temporal dimension is available. An overview of the token shift operation is given in \Cref{fig-token-shift-transformer}.

Either the patch tokens or the cls token - or both - can be shifted in the temporal dimension. The dimension that was shifted outside the tensor boundary is truncated and a padding zero is inserted.
The token shift operation shares temporal information between frames, therefore the temporal context can be better understood by the network. Other advantages of the TokShift operator are that it
requires zero parameters and zero FLOPs. It only shifts parts of the feature tensor. This removes the need of a spatio-temporal attention mechanism that is very computationally complex.

\subsection{Evo-ViT - Slow Fast Token Evolution}

Xu et al.~\cite{xu2022evo} propose a method for token dropping called slow-fast token evolution. It aims at reducing the number of parameters of the network by dropping tokens from
regions with low information density, e.g. the background tokens. They also introduce structure preserving token selection.
It utilizes informative tokens and placeholder tokens, the former of which are evolved in the token evolution stage. The third concept presented is global class attention, which evolves class attention
across layers of the network. 


Placeholder tokens do not contain useful information, opposite to informative tokens. Instead of preselecting uninformative tokens, the placeholder tokens are kept in the network training process
to keep the spatial structure of the network intact. Xu et al.~\cite{xu2022evo} observe that in the deeper layers of the network, informative tokens are assigned higher attention scores by the cls token.
In the slow-fast update scheme, representative tokens carry the information for the placeholder tokens. After the slow update of informative tokens, the representative tokens are used for a fast update of
placeholder tokens.

Global class attention enhances class attention by token evolution through the network.

\begin{align}
	A_{cls,g}^{k} = \alpha \cdot A_{cls}^{k-1} + (1 - \alpha)\cdot A_{cls}^{k},
\end{align}
where $A_{cls,g}^{k}$, $A_{cls}^{k}$ refer to global class attention and class attention in the k-th layer, respectively.
Global class attention is used to select placeholder and informative tokens.
Placeholder tokens are then summarized by representative tokens:
\begin{align}
	\mathbf{x}_{rep} = \phi_{agg}(\mathbf{x}_{ph}),
\end{align}
where $\phi_{agg} : \mathbb{R}^{(N - k) \times C} \rightarrow \mathbb{R}^{1 \times C}$ is an aggregation function - in this case the weighted sum.

These representative tokens are input to the transformer layer in tandem with the informative tokens. After the update step, the representative tokens
are used to update the placeholder tokens:

\begin{align}
	\mathbf{x}_{ph} \leftarrow \mathbf{x}_{ph} + \phi_{exp}(\mathbf{x}_{rep}^{(1)}) + \phi_{exp}(\mathbf{x}_{rep}^{(2)}).
\end{align}
$\phi_{exp} : \mathbb{R}^{1 \times C} \rightarrow \mathbb{R}^{(N - k) \times C}$ is an expanding function, e.g. a copy function.

This token update method reduces the redundancy in the tokens of the transformer network and accelerates inference times drastically with only small drops in accuracy.

\subsection{Efficient High-Order Spatial Interactions
with Recursive Gated Convolutions}

The Recursive Gated Convolution ($g^n$Conv) is introduced as a versatile module in the enhancement of vision Transformers and convolution-based models~\cite{rao2022hornet}. This novel operation incorporates gated convolutions and recursive designs, allowing for high-order spatial interactions. Notably, $g^n$Conv is flexible, customizable, and seamlessly integrates with different convolution variants. It extends two-order interactions in self-attention to arbitrary orders without introducing significant additional computation. In the domain of ViTs, the success is attributed to a spatial modeling paradigm involving input-adaptive, long-range, and high-order spatial interactions through self-attention. Although previous research has incorporated meta architectures~\cite{liu2022convnet}, input-adaptive weight generation~\cite{han2022connection}, and large-range modeling into CNN models~\cite{rao2021global}, a higher-order spatial interaction mechanism has been overlooked. The proposed $g^n$Conv efficiently addresses this gap by implementing the key ingredients in a convolution-based framework. Noteworthy properties of $g^n$Conv include efficiency, as its convolution-based implementation avoids the quadratic complexity of self-attention, and extendability, as it can achieve higher-order interactions with bounded complexity. Moreover, $g^n$Conv inherits translation equivariance from standard convolution, introducing beneficial inductive biases to major vision tasks and avoiding asymmetry associated with local attention. This module serves as a plug-and-play solution for enhancing the performance of various ViTs and convolution-based models.

\subsection{Token Sparsification for Faster Medical Image Segmentation}
Zhou et al. \cite{zhou2023token} present a token reduction method for medical image segmentation. Their proposed pipeline consists of the main steps: sparse encoding, token completion, and dense decoding. 
\\
In the first step, Soft topK Token Pruning modules (STP) are applied in between transformer blocks. Only the top K tokens are kept. The other tokens are pruned. To decide which tokens should be kept and which should be pruned, a score is estimated for each token. The score is estimated by a subnetwork $s_{\theta}$ that consists of two multi-layer perceptrons, average pooling, and a Sigmoid activation function. 
To sample the top K tokens, the scores are interpreted as a probability of the $i$th token ranking in the top K tokens. For $M_{i}=1$ the token is kept, for $M_{i}=0$ the token is pruned. To overcome the problem of a binary and therefore non-differentiable $M$, the function is approximated by $\Tilde{M}_{i}$. The formulas are:

\begin{equation}
    M_{i}= \underbrace{\mathbb{1}_{topK}\left(log\left(s_{i}\right) + g_{i}\right)}_\text{forward}
\end{equation} 
\begin{equation}   
    \underbrace{\Tilde{M_{i}} = \frac{exp\left(\left(log\left(s_{i}\right)+g_{i}\right)/\tau\right)}{\Sigma_{j=1}^{n}exp\left(\left(log\left(s_{j}\right)+g_{j}\right)/\tau\right)}}_\text{backward}
\end{equation}

Where $g_{i}$ is the Gumbel Softmax \cite{gumbel1954statistical}. During inference, the top tokens are selected without the added Gumbel noise.

In the second step, the sparse tokens are completed to generate a dense output later. The pruned tokens $\{\bar{\mathbf{z}}_{1}, \bar{\mathbf{z}}_{2}, \bar{\mathbf{z}}_{3}\}$ from each layer are added with learnable block tokens and concatenated with the final output tokens $\bar{\mathbf{z}}_{L}$. The tokens are then rearranged to their original spatial order and sine-cosine position embeddings are added. Finally, the tokens are used as input for a transformer block. 

For the third step, the dense decoding and the generation of the segmentation output the decoder of UNETR \cite{hatamizadeh2022unetr} is used. This token sparsification method allows a token reduction of up to $90\%$ and a highly increased throughput while keeping the accuracy the same.


\subsection{Vision Transformer with Bi-Level Routing Attention}

Zhu et al. present a pioneering ViT, referred to as BiFormer~\cite{zhu2023biformer}, which puts forth a dynamic sparse attention mechanism through a bi-level routing strategy. The primary objective is to advance computational efficiency while prioritizing content awareness. The key proposition involves empowering each query to selectively attend to a restricted subset of the most semantically \textit{relevant} key-value pairs. To achieve global attention with optimal efficiency, the authors advocate for a region-to-region routing approach. Rather than filtering out irrelevant key-value pairs at the token level, a coarse-grained region-level affinity graph is constructed, and subsequent pruning retains only the top-\textit{k} connections for each node.

\begin{figure}[]
    \centering
    \includegraphics[width=\columnwidth]{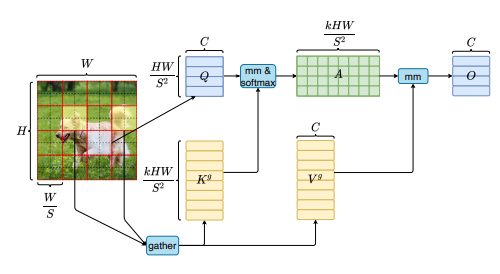}
    \caption{Architecture of the Bi-Level Routing Attention. From \cite{zhu2023biformer}.}
    \label{fig-Biformer}
\end{figure}

In this paradigm, each region is tasked with attending solely to the top-\textit{k} routed regions, streamlining the attention mechanism. The subsequent step involves token-to-token attention, a non-trivial task given the spatial scattering of key-value pairs. In contrast to conventional sparse matrix multiplication, which proves inefficient on modern GPUs, the proposed solution involves gathering key/value tokens to engage in hardware-friendly dense matrix multiplications. This innovative approach, termed Bi-level Routing Attention (BRA), integrates a region-level routing step and a token-level attention step. The concept is demonstrated in \Cref{fig-Biformer}

In comparison to static patterns of sparse attention, BiFormer incorporates an additional step to identify the regions to attend. This entails constructing and pruning a region-level graph and gathering key-value pairs from the routed regions. While this step operates at a coarse region level and does not significantly increase the computational load, it introduces extra GPU kernel launches and memory transactions. Consequently, despite comparable FLOPs on GPU, BiFormer exhibits lower throughput than some existing models due to the overheads associated with kernel launch and memory bottlenecks.

\section{Other Transformer Architectures} \label{other}

Lastly, methods are present that improve another aspect of the transformer that does not fit into the previous categories.

\subsection{FocalNet - Focal Modulation Networks}

Yang et al. introduce the Focal Modulation Network~\cite{yang2022focal}, a network that replaces self-attention with focal modulation.

Like in self-attention, a query token is computed from the input feature.
Instead of calculating the attention scores first and multiplying them with the values (summarized as \textit{interaction} in~\cite{yang2022focal}) and aggregating the resulting context vectors,
focal modulation first aggregates the context features to then compute the interaction:

\begin{align}
	\mathbf{y}_{i} = \mathit{T}_{2}(\mathit{M}_{2}(\mathbf{x}_{i}, \mathbf{X}),\mathbf{x}_{i})
\end{align}
Aggregation starts with \textit{hierarchical contextualization}, computing local or global context for fine or coarse-grained features. The features are aggregated into one feature vector
with \textit{gated aggregation}. This feature vector is referred to as the \textit{modulator}.

Hierarchical Contextualization first projects the input feature to a new feature space. Afterwards, L depth-wise convolutions are used:

\begin{align}
	\mathbf{Z}^{l} = f_{a}^{l}(\mathbf{Z}^{l-1}) \triangleq \operatorname{GELU}(\mathdash{DW-Conv}(\mathbf{Z}^{l-1})).
\end{align}
$f_{a}^{l}$ is the contextualization function at the $l$-th level. In each focal level $l$, a gated aggregation is computed:

\begin{align}
	\mathbf{Z}^{out} = \sum_{l=1}^{L+1} \mathbf{G}^{l} \odot \mathbf{Z}^{l},
\end{align}
where $G^{l}$ are the spatial- and level-aware gating weights for level $l$ - the weights are obtained through a linear layer.
The modulator is another linear layer $\mathbf{M} = h(\mathbf{Z}^{out}) \in \mathrm{R}^{H \times W \times C}$.

The main advantage of focal modulation is an improvement of computational complexity over self-attention. Similar to efficient attention,
the order of aggregation and interaction are exchanged to not result in a quadratic complexity relative to the number of tokens.

The total time complexity to compute a feature map with focal modulation is $O(HW \times (3C^{2} + C(2L + 3) + C\sum_{l}(k^{l})^{2}))$.
According to~\cite{yang2022focal}, Swin-T windowed attention with a window size of w has a complexity of $O(HW \times (3C^{2} + 2Cw^{2}))$. $L$ refers to the
number of depthwise convolution layers and $k^{l}$ to the kernel size of said convolution. Focal modulation is more efficient because $L$ and $(k^{l})^{2}$
are usually much smaller than C.



\subsection{DeepViT - Deep Vision Transformer}

Zhou et al.~\cite{zhou2021deepvit} analyze the effect of increasing the depth of transformer networks. Unlike in CNNs, where increasing the depth increases the richness of the feature representations and thus
the performance increases, increasing the depth of the standard ViT actually stagnates the performance and it drops when the depth is increased further.

In order to understand this phenomenon, they calculate the cross-layer similarity at each transformer layer and observe that in deeper layers, the attention maps of different heads become more similar
to each other. This is called attention collapse and it prevents deeper networks from learning more context than shallower ones.

In order to solve the aforementioned problem, re-attention is introduced. The attention maps from the different attention heads are aggregated before multiplying by the values:

\begin{align}
	\mathdash{Re-Attention}(\mathbf{Q}, \mathbf{K}, \mathbf{V}) = \operatorname{LN}(\Theta^{T}(\operatorname{softmax}(\frac{\mathbf{QK}^{T}}{\sqrt{d}})))\mathbf{V}.
\end{align}
$\Theta$ is a transformation matrix which is multiplied by $\mathbf{A}$ along the head dimension ($\mathbf{A}$ refers to the attention map).
This method works because the difference between attention maps in different heads is usually quite high, which leads to a more diverse output feature map.

The re-attention mechanism is employed instead of the self-attention in the standard ViT architecture. Zhou et al. show that their model's performance increases monotonically with model depth. This enables future architectures
to scale their networks to larger depths.

\subsection{LeViT}

LeViT by Graham et al.~\cite{graham2021levit} combines CNN and transformer, aiming to reduce the inference time of the network. 

ResNet-50~\cite{resnet} is combined with the ViT architecture based on DeiT~\cite{touvron2021training}. The patch embedding is done via 4 layers of $3 \times 3$ convolutions instead of one $16 \times 16$ convolution
to reduce computation time. The classification token is removed and instead average pooling on the last activation map is utilized to produce a classification feature.

To adapt the attention to the CNN architecture, LeViT attention uses $1 \times 1$ convolutions to compute keys, queries, and tokens for the attention mechanism. 
Instead of max-pooling operations, shrink attention is employed between each stage which reduces the size of the activation map by 1/2.

Positional encoding is replaced by attention bias:

\begin{align}
	A^{h}_{(x,y),(x',y')}=\mathbf{Q}_{(x,y),:} \cdot \mathbf{K}_{(x',y'),:} + \mathbf{B}^{h}_{|x-x'|,|y-y'|}.
\end{align}
The first term is standard self-attention. The second term is the attention bias. Attention bias is translation-invariant.
The resulting value $A^{h}$ is the attention value between two pixels
$(x,y)$ and $(x',y')$. This bias term allows the model to train with flip invariance.

The MLP blocks of the ResNet-50 are also reduced in size. In LeViT, one MLP block consists of an expansion by a factor 2, a $1 \times 1$ convolution, batch normalization, and reduction by a factor 2 (which is 4 in standard ResNet-50).
This makes the attention block and MLP blocks use approximately the same number of FLOPs.

The LeViT architecture is one approach how to combine the transformer architecture with convolutional architectures. It optimizes the inference times of the transformer without regard to the number of parameters.
It matches state-of-the-art approaches in performance while increasing the inference speeds - through increasing the number of parameters compared to similarly performant networks.

\subsection{CvT - Convolutional Vision Transformer}

Wu et al.~\cite{wu2021cvt} propose another architecture that integrates convolutions into the transformer - the Convolutional Vision Transformer (CvT). 

CvT introduces convolutions in two parts of the hierarchical transformer architecture - convolutional token embedding and the convolutional projection layer.
Convolutional token embedding models local spatial context by convolutions on overlapping patches. This reduces the number of tokens in each stage
while increasing the feature dimension. The result of each convolution is a token, and the series of resulting tokens is fed to a stack of convolutional transformer blocks.

The convolutional transformer block consists of multi-head self-attention as in the ViT~\cite{dosovitskiy2020vit}, but the projection is a convolution instead of a linear layer. 

Convolutional projection allows an additional step where local context can be modeled implicitly through the convolution operation. It also enables the network to reduce the sizes of the $\mathbf{K}$ and $\mathbf{V}$ matrices, reducing the computational complexity.
If both are the same size, the output also stays the same size.
First, the token sequence is reshaped into a 2D token map. Convolutional projection applies a set of $s \times s$ convolutions to the token map, and depending on the stride the size of the output token map may be reduced.
The output token map is then flattened again to receive the queries, keys and values as input to the multi-head self-attention. Squeezed convolutional projection applies stride 2 to the convolution for keys and values, which
reduces the size of the respective tensors. This reduces the performance only minimally as neighbouring pixels usually contain redundant information. It decreases the cost of the self-attention operation by a factor of 4,
which is drastic given the quadratic complexity of it.
As a side note, the linear projection layer of the ViT could be implemented as a set of $1 \times 1$ convolutions, which makes convolutional projection a generalization of linear projection.


The benefits of including convolutions into the transformer model are: Local context is implicitly modeled by the convolution operation, shared weights make the method more efficient. It also keeps the
advantages of the transformer architecture: Modeling of global context and good generalization capabilities.

\subsection{Vision Transformer with Deformable Attention}

\begin{figure}
	\centering
	\includegraphics[width=\columnwidth]{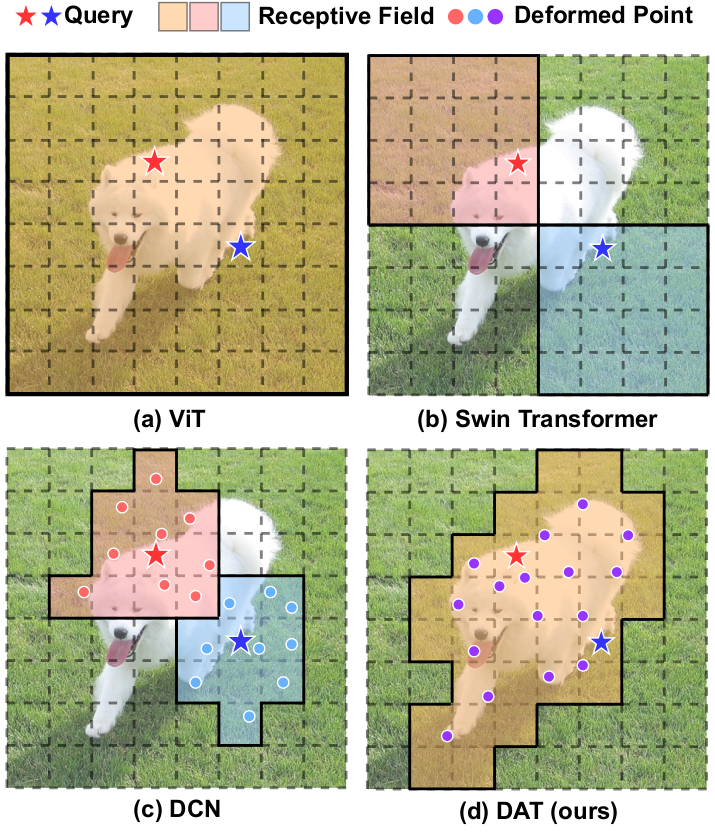}
	\caption{Standard self-attention in~\cite{dosovitskiy2020vit}(a). Windowed self-attention in~\cite{liu2021swin} (b). DCN in (c)~\cite{dai2017deformable}. Deformable Queries in (d). From~\cite{xia2022vision}.}\label{fig-deformable-attention}
\end{figure}

Another method of computing queries is presented by  Xia et al.~\cite{xia2022vision}. They propose the Vision Transformer with deformable attention. As the name suggests,
attention is not calculated on static patches of the same size. Instead, deformed points determine the queries. The concept is shown in~\Cref{fig-deformable-attention}.

Queries are calculated from the input feature, but keys and values are calculated from a set of deformed points.
First, reference points are set at equidistant positions in the input. An offset network deforms the points depending on the structure of the input feature. Value and key patches are computed based on these deformable points.

The deformable points are more closely related to the structure of the input feature, unlike arbitrarily sampled patches. 

The cost of the DMHA module is linear with regards to the channel dimension, which is minor relative to the quadratic complexity of self attention.


\subsection{VAN - Visual Attention Network}
\begin{figure}[]
    \centering
    \includegraphics[width=\columnwidth]{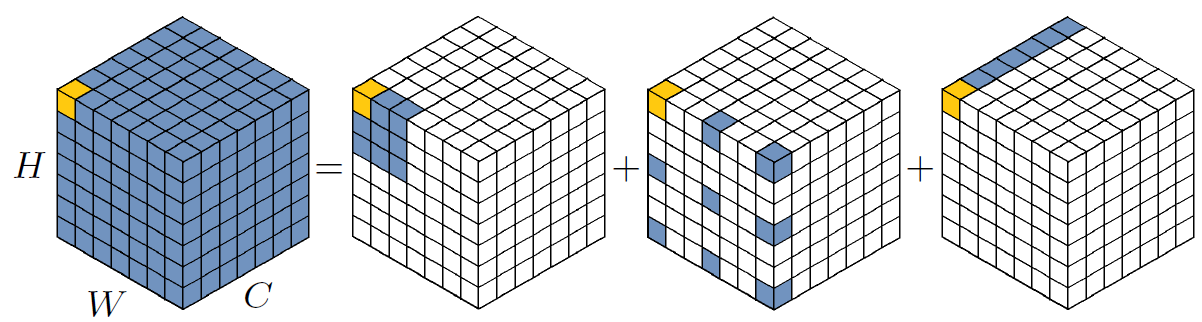}
    \caption{A large convolution kernel is constructed from three small convolutions. From \cite{guo2022attention}.}
    \label{fig-visual_attention}
\end{figure}

A different kind of attention, named Visual Attention, is proposed by Guo et al. \cite{Guo2023}. Self-attention has three major drawbacks: 1. It treats images as 1D sequences and ignores their 2D structure, which provides important information. 2. The quadratic complexity with respect to the number of tokens, limits the input size of the images. 3. The channel adaptability is ignored. Visual attention tries to overcome these shortcomings.\\
The main idea of the attention mechanism is to produce an attention map, that highlights important parts and neglects unimportant ones. Self-attention is one possibility to create these attention maps. Another possibility is utilizing large kernel convolution. The drawbacks of these large kernel convolutions are the high number of parameters and computational cost. The authors overcome these issues by constructing a large kernel by decomposing it into three smaller convolution operations, shown in figure \ref{fig-visual_attention}. A $K \times K$ convolution can be divided into a $\left(2d-1\right)\times \left(2d-1\right)$ depth-wise spatial convolution for local attention, a $\left\lceil\frac{K}{d}\right\rceil \times \left\lceil\frac{K}{d}\right\rceil$ dilated depth-wise convolution for global context and a $1\times 1$ convolution to incorporate channel information. By using depth-wise convolutions a large kernel is constructed with a low number of parameters and small computation costs.

The formulas to create the attention map with the Large Kernel Attention (LKA) and the output feature map are:
\begin{equation}
    A = Conv_{1 \times 1} \left(DWD Conv\left(DW Conv\left(F\right)\right)\right),
\end{equation}
\begin{equation}
    Output = A \otimes F.
\end{equation}
The importance of features is denoted in the attention map $Attention \in \mathcal{R}^{C\times H \times W}$. The $Output$ is created by the element-wise multiplication of the input features $F\in \mathcal{R}^{C\times H\times W}$. LKA combines a local receptive field with global information and channel information. The complexity scales linearly with the input size. \\
The authors propose a new network architecture called Visual Attention Network (VAN). Here, LKA is used as the main building block of the network. The network is trained for various tasks, including classification, object detection, and semantic segmentation.\\
Large kernel attention combines the advantages of self-attention and convolutions. Global information and local features are captured within a single block. The linear complexity makes it feasible for large inputs.

\subsection{Medical Transformer: Gated Axial-Attention for Medical Image Segmentation}
Valnarasu et al. \cite{valanarasu2021medical} propose a position-sensitive gated attention mechanism and a local-global training strategy. Medical Image datasets are often small and it is therefore crucial to develop networks, that converge on a small dataset. Positional encodings are important due to the loss of position information in transformer networks. However, positional encodings may not be accurate enough when trained on small-scale datasets. Therefore, the authors introduce a gating mechanism to control the influence of the positional bias. The learnable gating parameters assign a high weight to accurately learned encoding and a value close to zero otherwise. The gated axial attention mechanism for the width axis can be expressed by: 
\begin{multline}
    y_{ij} = \sum_{w=1}^{W} (q_{ij}^Tk_{iw}+G_Qq_{ij}^Tr_{iw}^k + G_Kk_{iw}^Tr_{iw}^k) \\
    \times (G_{V1}v_{iw} + G_{V2}r_{iw}^v),
\end{multline}
with the learnable gating parameters $G_Q, G_K, G_{V1}, G_{V2}$ and positional encodings $r_{iw}^q, r_{iw}^k, r_{iw}^v$.
Furthermore, the authors introduce a Local-Global training strategy. A global branch operates on patches of the original image resolution and a local branch on partial image patches. The features of both branches are fused by addition and a convolution layer. While the global branch captures important global dependencies, the local branch can focus on fine details.

\subsection{D-LKA-Net - Deformable Large Kernel Attention}
\begin{figure}[]
    \centering
    \includegraphics[width=\columnwidth]{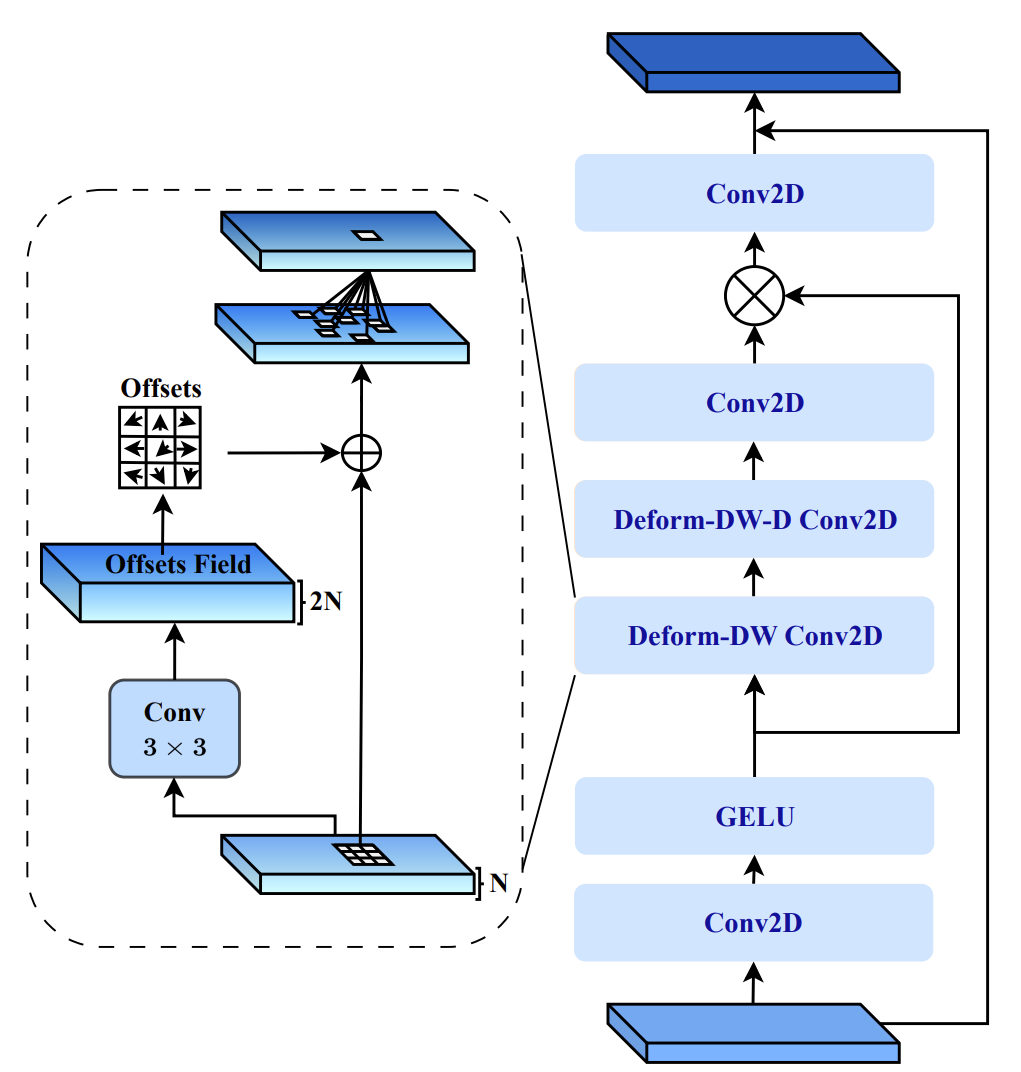}
    \caption{Architecture of the deformable LKA module. From \cite{azad2023selfattention}.}
    \label{fig-deformable-lka}
\end{figure}

Azad et al. \cite{azad2023selfattention} propose a novel attention mechanism that combines LKA \cite{Guo2023} and deformable convolutions \cite{dai2017deformable}. An attention map is constructed by deformable large kernel convolutions. To improve the efficiency, the large deformable convolution kernel is created from smaller deformable convolutions. This allows the network to learn an adaptive deformation grid with adjusted receptive fields for each input. The 2D deformable LKA module is presented in Figure \ref{fig-deformable-lka}.
The computational complexity is determined by the kernel size of the deformable convolutions and the image size, as well as the channels:
\begin{multline}
    FLOPs = C(C + 2K_{DDW}^4 + K_{DDW}^2(1+C) + 2K_{DW}^4 \\
    + K_{DW}^2(1+C))\times HW,
\end{multline}
with channels $C$, dilated depth-wise kernel size $K_{DDW}$, depth-wise kernel size $K_{DW}$, height $H$ and width $W$. The computational complexity is linear with respect to the image size and quadratic with respect to the number of channels.
Deformable LKA captures spatial and channels information in an adaptive manner while remaining efficient in terms of parameters and computations.

\begin{table*}[!th]
	\centering
	\caption{A brief description of the reviewed enhanced efficiency in Vision Transformer Networks.}
	\label{tab-transformer-architectures-1}
	\resizebox{\textwidth}{!}{
		\begin{tabular}{ p{2.5cm} | p{1cm} | p{19cm} | p{0.6cm}} 
				\hline
				\rowcolor{gray!20}
				\textbf{Methods} 
				& \textbf{Params} 
				& \textbf{Highlights}
				& \textbf{Year}
				\\ \hline 
			
				\rowcolor{gray!5}
				CrossViT-18~\cite{chen2021crossvit}
				& \adjustbox{valign=b}{44.3 M}
				& $\bullet$ A dual branch transformer is proposed with large and small patch sizes. This enables different transformer depths for both branches.  $\bullet$ To fuse the respective tokens, token cross attention is proposed. It attends cls tokens of one branch with patch tokens of the other.  $\bullet$ The model complexity is reduced because the small branch does not contain as many transformer blocks.  $\bullet$ A multiscale representation of the input is achieved. $\bullet$ Cross token attention works only for classification tasks. 
				& 2021
				\\ \hline 

				\rowcolor{yellow!10}
				Swin-B~\cite{liu2021swin} 
				& 88 M 
				& $\bullet$ A hierarchical shifted-window transformer is proposed.  $\bullet$ Shifted-window attention captures long-range context via two consecutive attention operations on shifted windows. $\bullet$ The computational complexity is reduced compared to standard ViT.  $\bullet$ Global context is captured.  $\bullet$ Requires pretraining on another dataset 
				& 2022 
				\\ \hline

				\rowcolor{gray!5}
				XCiT-S24~\cite{ali2021xcit} 
				& 48 M 
				& $\bullet$ They propose the cross-covariance transformer. $\bullet$ Cross-covariance attention, a form of channel attention, is introduced.  $\bullet$ The computational complexity of the attention mechanism is reduced to linear with regards to the number of tokens $N$. $\bullet$ The cross-covariance attention does not capture spatial context explicitly.
				& 2021 
				\\ \hline

				\rowcolor{yellow!10}
				DaViT-Base~\cite{ding2022davit} 
				& 87.9 M 
				& {$\bullet$ A dual attention - spatial window self attention followed by channel group attention is proposed.  $\bullet$ The windowed attention reduces the complexity while the channel attention learns the global context.  $\bullet$ Regular self attention is used in the windowed attention, which has quadratic complexity with regards to $N$.} 
				& 2022 
				\\ \hline

				\rowcolor{gray!5}
				MSG-B~\cite{fang2022msg} 
				& 84 M 
				& {$\bullet$ They propose the MSG token.  $\bullet$ The shuffle operation exchanges information between local windows.  $\bullet$ The quadratic complexity of self attention is kept to local windows.  $\bullet$ Regular self attention is used in the windowed attention, which has quadratic complexity with regards to $N$.  $\bullet$ MSG Tokens cannot be ported easily to other architectures.} 
				& 2022 
				\\ \hline

				\rowcolor{yellow!10}
				MISSFormer-B~\cite{huang2022missformer} 
				& 42.5 M 
				& {$\bullet$ The MISSFormer architecture is introduced. It utilizes efficient self attention and the enhanced transformer context bridge.  $\bullet$ Efficient self attention reduces the spatial dimension by a reduction ratio $R$.  $\bullet$ The enhanced transformer context bridge concatenates outputs of the skip connections and performs efficient attention.  $\bullet$ The computational complexity of the self attention is reduced by the factor $R$.  $\bullet$ The model can be trained from scratch - it does not require pretraining.  $\bullet$ Regular self attention is used in the windowed attention, which has quadratic complexity with regards to $N$.} 
				& 2022 
				\\ \hline

				\rowcolor{gray!5}
				RegionViT-B~\cite{chen2021regionvit} 
				& 72.7 M 
				& {$\bullet$ Regional-to-local attention is proposed.  $\bullet$ The computational complexity of the attention is reduced by $O(N/M)$ where $N$ is the number of tokens and $M$ is the window size.  $\bullet$ Global context is learned by regional tokens.  $\bullet$ Regular self attention is used in the windowed attention, which has quadratic complexity with regards to $N$.} 
				& 2022 
				\\ \hline

				\rowcolor{yellow!10}
				nnFormer \cite{zhou2023nnformer} 
				& 150 M 
				& {$\bullet$ Mix of Local and Global self-attention.  $\bullet$ Cross attention in the skip connections is proposed.  $\bullet$ Full self-attention in bottleneck.  $\bullet$ The network has a lot of parameters and a large number of FLOPs.  $\bullet$ Window attention in the encoder and decoder is limiting the receptive field size.} 
				& 2023 
				\\ \hline

				\rowcolor{gray!5}
				EdgeNeXt-S~\cite{maaz2022edgenext} 
				& 5.6 M 
				& {$\bullet$ A transformer for edge devices is implemented.  $\bullet$ Transpose attention is proposed.  $\bullet$ Transpose attention is linear with regards to $N$.  $\bullet$ Transpose attention does not capture spatial context explicitly.} 
				& 2023 
				\\ \hline

				\rowcolor{yellow!10}
				GCViT-B~\cite{hatamizadeh2023global} 
				& 90 M 
				& {$\bullet$ The global query generator is presented.  $\bullet$ Global MSA is proposed. It utilizes local values and keys, and the global query.  $\bullet$ Global context is captured by the global queries.  $\bullet$ Regular self attention is used for local attention, which has quadratic complexity with regards to $N$.} 
				& 2023 
				\\ \hline

				\rowcolor{gray!5}
				FasterViT-2~\cite{hatamizadeh2023fastervit} 
				& 75.9 M 
				&{$\bullet$ FasterViT optimizes image throughput by combining fast local representation learning from CNNs and global modeling from ViTs.  $\bullet$ Hierarchical Attention (HAT) in FasterViT efficiently reduces computational costs using window-based self-attention with dedicated carrier tokens (CTs).  $\bullet$ CTs alternate between sub-global and local self-attention, forming hierarchical attention for comprehensive information processing.} 
				& 2023 
				\\ \hline

				\rowcolor{yellow!10}
				EffFormer-B\footnotemark~\cite{shen2021efficient} 
				& 22.3 M 
				& {$\bullet$ Efficient attention is proposed.  $\bullet$ The complexity of the attention mechanism is reduced to linear with regards to $N$.  $\bullet$ The spatial context is captured.} 
				& 2021 
				\\ \hline

				\rowcolor{gray!5}
				FocalNet-B~\cite{yang2022focal} 
				& 88.7 M 
				& {$\bullet$ Focal modulation is presented.  $\bullet$ Hierarchical contextualization is proposed. It is applied in the focal modulation module after obtaining a representation of the input.  $\bullet$ Focal modulation first aggregates context vectors which reduces the redundancy of the model.  $\bullet$ Hierarchical contextualization gains local context by consecutive CNNs.  $\bullet$ Focal modulation is not an attention mechanism, but couples principles of attention and convolution.} 
				& 2022 
				\\ \hline

				\rowcolor{yellow!10}
				Spectral-Spatial Net~\cite{sun2022fusing} 
				& - 
				& {$\bullet$ A multi-fusion architecture is used.  $\bullet$ Transformer is applied to hyperspectral (HSR) images.  $\bullet$ Multi-fusion shows that concatenation fusion has the highest classification score.  $\bullet$ Can be applied only to HSR image classification.} 
				& 2022 
				\\ \hline

				\rowcolor{gray!5}
				SCViT-B~\cite{lv2022scvit} 
				& 22.1M 
				& {$\bullet$ A spatial-channel transformer is presented.  $\bullet$ Channel information is considered for HSR imagery.  $\bullet$ The lightweight channel attention (LCA) module weighs channel information of the classification token, increasing the classification performance.  $\bullet$ The method is only applicable to classification problems.} 
				& 2022 
				\\ \hline

				\rowcolor{yellow!10}
				CAA~\cite{huang2022channelized} 
				& - 
				& {$\bullet$ Channelized Axial Attention is proposed.  $\bullet$ Channel and spatial attention are combined within one module.  $\bullet$ Axial attention splits row and column attention.} 
				& 2022 
				\\ \hline

				\rowcolor{gray!5}
				{Semantic-enhanced  Dual Attention~\cite{ma2022knowing}} 
				& 33.8 M 
				& {$\bullet$ A transformer architecture for image captioning is presented.  $\bullet$ Semantic-enhanced dual attention is utilized.  $\bullet$ The presented dual attention is independent of the image caption - it can be used in other architectures.  $\bullet$ The model is very small considering it uses dual attention.  $\bullet$ The architecture is applied to image captioning.  $\bullet$ Faster-RCNN is employed, making this architecture not a pure transformer.} 
				& 2022 
				\\ \hline

				\rowcolor{yellow!10}
				UNETR++~\cite{shaker2022unetr++} 
				& 42.96 M 
				& {$\bullet$ Efficient paired attention is presented.  $\bullet$ Weight sharing reduces number of parameters.  $\bullet$ Spatial and channel attention is combined.  $\bullet$ The 3D data structure is neglected.} 
				& 2022 
				\\ \hline

				\rowcolor{gray!5}
				DeepViT-32B~\cite{zhou2021deepvit} 
				& 48.1 M 
				& {$\bullet$ The Deep Vision Transformer is proposed.  $\bullet$ Re-Attention, which enables deeper transformer architectures, is presented.  $\bullet$ Stacking more transformer blocks does not saturate the performance - it monotonically increases.  $\bullet$ The self attention is still quadratic in complexity} 
				& 2021 
				\\ \hline

				\rowcolor{yellow!10}
				LeViT-384~\cite{graham2021levit} 
				& 39.1 M 
				& {$\bullet$ A transformer model applying convolutions is proposed.  $\bullet$ $^{1}$ The inference is much faster than for pure attention transformers.  $\bullet$ The model size is comparable to equally performant models  $\bullet$ The model relies on convolutions} 
				& 2021 
				\\ \hline

				\rowcolor{gray!5}
				CvT-21~\cite{wu2021cvt} 
				& 31.5 M 
				& {$\bullet$ A transformer architecture based on convolutions is proposed. $\bullet$ Convolutional transformer block and convolutional projection modules are introduced. $\bullet$ The model requires less training data, similar to CNNs, to perform well. $\bullet$ Convolutional projection captures more local context than linear projection. $\bullet$ The model relies on convolutions.} 
				& 2021 
				\\ \hline

				\rowcolor{yellow!10}
				DAT-B~\cite{xia2022vision} 
				& 88 M 
				& {$\bullet$ Deformable attention is presented. $\bullet$ Relevant keys and values are adapted to the input, whereas irrelevant background patches have less importance. $\bullet$ The performance is improved while keeping important information. $\bullet$ The deformable attention is difficult to integrate due to the offset network.} 
				& 2022 
				\\ \hline

				\rowcolor{gray!5}
				DynamicViT-LV-M/0.8~\cite{rao2021dynamicvit} 
				& 57.1 M 
				& {$\bullet$ A dynamic token sparsification network is proposed. $\bullet$ Throughput and model complexity are greatly reduced. $\bullet$ Performance only drops slightly compared to similarly-sized models. $\bullet$ The model requires Gumbel-softmax training because it is non-differentiable.} 
				& 2021 
				\\ \hline

				\rowcolor{yellow!10}
				LV-ViT-M~\cite{jiang2021all} 
				& 56 M 
				& {$\bullet$ A network is proposed that utilizes information from all tokens for classification. $\bullet$ Performance is increased. $\bullet$ Only a minor increase in complexity is required. $\bullet$ The method is applicable to downstream tasks like segmentation.} 
				& 2021 
				\\ \hline

				\rowcolor{gray!5}
				Bi-Former-B~\cite{zhu2023biformer}
				& 58 
				& {$\bullet$ Biformer enhances efficiency with dynamic sparse attention via Bi-level Routing Attention (BRA). $\bullet$ The region-to-region routing strategy streamlines global attention, allowing queries to focus on relevant key-value pairs. $\bullet$ BRA optimizes efficiency with hardware-friendly dense matrix multiplications, overcoming GPU inefficiencies. $\bullet$ Biformer exhibits lower GPU throughput due to introduced overheads, including extra kernel launches and memory transactions during region-level graph construction and pruning.} 
				& 2023 
				\\ \hline

				\rowcolor{yellow!10}
				TokShift (MR)~\cite{zhang2021token} 
				& 85.9 M 
				& {$\bullet$ A token shift operation is introduced. $\bullet$ Token shift requires zero additional parameters.  $\bullet$ Video data can be processed by the network  $\bullet$ The method is only applicable to video data for all shift variants.} 
				& 2021 
				\\ \hline

				\rowcolor{gray!5}
				Evo-LeViT-384~\cite{xu2022evo} 
				& 39.6 M 
				& {$\bullet$ Slow-fast token evolution is proposed. $\bullet$ Long-range dependencies between tokens are more efficiently modeled.  $\bullet$ Instead of token pruning, uninformative tokens are reduced in size but kept through the hierarchy. $\bullet$ The token evolution is difficult to integrate into any network.  $\bullet$ The standard self attention is used, which is quadratic in complexity.} 
				& 2022 
				\\ \hline

				\rowcolor{yellow!10}
				Token sparsification \cite{zhou2023token} 
				& - 
				& {$\bullet$ Token scores are estimated via a small sub-network.  $\bullet$ Model complexity is reduced, throughput is increased.  $\bullet$ Performance drops only by a small amount. $\bullet$ Unused tokens are later restored for the decoder. $\bullet$ Only the encoder is changed.  $\bullet$ Gumbel softmax is needed.} 
				& 2023 
				\\ \hline

				\rowcolor{gray!5}
				VAN (B5) \cite{Guo2023} 
				& 90M 
				& {$\bullet$ Visual Attention is proposed.  $\bullet$ Visual Attention is efficient.  $\bullet$ Operates in channel and spatial domain.  $\bullet$ Integration into existing work is easy.} 
				& 2023 
				\\ \hline

				\rowcolor{yellow!10}
				MedT \cite{valanarasu2021medical} 
				& - 
				& {$\bullet$ A gating mechanism for axial attention is presented.  $\bullet$ Computation reduction due to the axial attention mechanism  $\bullet$ The local-global strategies captures both local and global features. $\bullet$ Axial attention fails to capture spatial relations.} 
				& 2021 
				\\ \hline

				\rowcolor{gray!5}
				D-LKA~\cite{azad2023selfattention} 
				& 101.64 
				& {$\bullet$ DLKA combines LKA and deformable convolutions for a novel attention mechanism.  $\bullet$ The attention map is efficiently constructed using deformable large kernel convolutions, enabling adaptive deformation grids for each input.  $\bullet$ The 2D deformable LKA module captures spatial and channel information with linear complexity in image size and quadratic complexity in the number of channels, maintaining efficiency.} 
				& 2023 
        		\\ \hline

		\end{tabular}
	}
\footnotetext{text} 
\end{table*}

\section{Discussion} 
\label{sec:discuss}

\begin{figure*}
    \centering
	\includegraphics[width=0.87\textwidth]{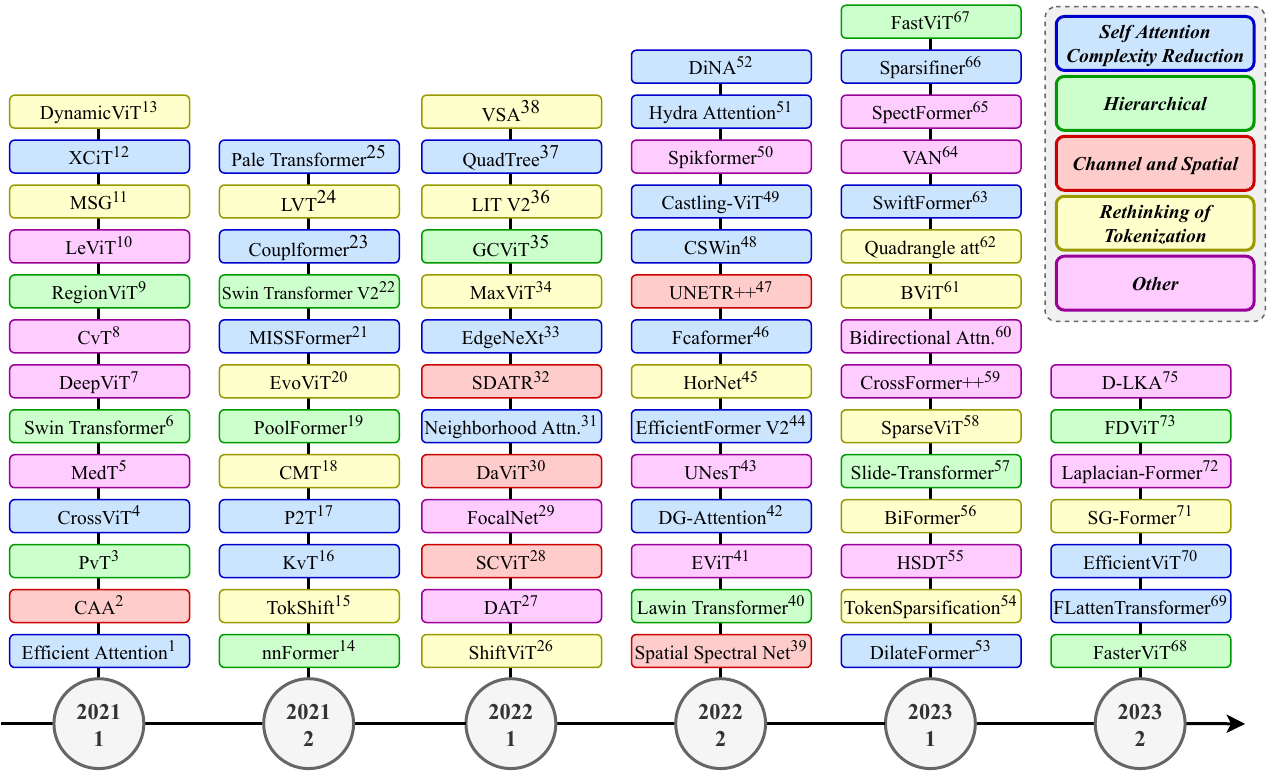}
	\caption{A timeline of the contributions to transformer architectures.
     1:~\cite{shen2021efficient}, 2:~\cite{huang2022channelized}, 3:~\cite{wang2022pvt},
     4:~\cite{chen2021crossvit}, 5:~\cite{valanarasu2021medical}, 6:~\cite{liu2021swin},
     7:~\cite{zhou2021deepvit}, 8:~\cite{wu2021cvt}, 9:~\cite{chen2021regionvit},
     10:~\cite{graham2021levit}, 11:~\cite{fang2022msg}, 12:~\cite{ali2021xcit},
     13:~\cite{rao2021dynamicvit}, 14:~\cite{zhou2023nnformer}, 15:~\cite{zhang2021token},
     16:~\cite{wang2022kvt}, 17:~\cite{wu2022p2t}, 18:~\cite{guo2022cmt},
     19:~\cite{yu2022metaformer}, 20:~\cite{xu2022evo}, 21:~\cite{huang2022missformer},
     22:~\cite{liu2021swinv2}, 23:~\cite{lan2021couplformerrethinking}, 24:~\cite{yang2021lite},
     25:~\cite{wu2021pale}, 26:~\cite{wang2022shift}, 27:~\cite{xia2022vision},
     28:~\cite{lv2022scvit}, 29:~\cite{yang2022focal}, 30:~\cite{ding2022davit},
     31:~\cite{hassani2023neighborhood}, 32:~\cite{ma2022knowing}, 33:~\cite{maaz2022edgenext},
     34:~\cite{tu2022maxvit}, 35:~\cite{hatamizadeh2023global}, 36:~\cite{pan2023fast},
     37:~\cite{tang2022quadtree}, 38:~\cite{zhang2023vsa}, 39:~\cite{sun2022fusing},
     40:~\cite{yan2023lawin}, 41:~\cite{feng2022evit}, 42:~\cite{liu2022dynamic},
     43:~\cite{yu2022unest}, 44:~\cite{li2023rethinking}, 45:~\cite{rao2022hornet},
     46:~\cite{zhang2023fcaformer}, 47:~\cite{shaker2022unetr++}, 48:~\cite{dong2022cswin},
     49:~\cite{you2023castlingvit}, 50:~\cite{zhou2022spikformer}, 51:~\cite{bolya2022hydra},
     52:~\cite{hassani2023dilated}, 53:~\cite{jiao2023dilateformer}, 54:~\cite{zhou2023token},
     55:~\cite{lai2023hybrid}, 56:~\cite{zhu2023biformer}, 57:~\cite{pan2023slide},
     58:~\cite{chen2023sparsevit}, 59:~\cite{wang2023crossformer}, 60:~\cite{fan2023lightweight},
     61:~\cite{li2023bvit}, 62:~\cite{zhang2023quad}, 63:~\cite{shaker2023swiftformer},
     64:~\cite{guo2022visual}, 65:~\cite{patro2023spectformer}, 66:~\cite{wei2023sparsifiner},
     67:~\cite{vasu2023fastvit}, 68:~\cite{hatamizadeh2023fastervit}, 69:~\cite{han2023flatten},
     70:~\cite{liu2023efficientvit}, 71:~\cite{Ren_2023_ICCV},72:~\cite{azad2023laplacianformer},
     73:~\cite{Xu_2023_FDViT}, 74:~\cite{azad2023selfattention}.
    }\label{fig-timeline}
\end{figure*}

In this survey, we have systematically explored recent advancements in enhancing the efficiency of ViT models within the domain of CV. The pivotal role of attention mechanisms in the development of ViTs cannot be overstated, given their demonstrated ability to significantly boost model performance across various vision tasks in diverse research fields. Our survey introduces a thorough taxonomy in~\Cref{fig-taxonomy-design} designed to categorize and elucidate the multitude of attention mechanisms, strategically organized based on their impact on the redesign of ViTs for efficiency enhancement. To further facilitate understanding, we present a detailed comparison in~\Cref{tab-transformer-architectures-1} and~\Cref{tab-attention-complexity}, offering insights into key aspects such as contributions, trainable parameters, FLOPS, MACs, time complexity, and issue date. Additionally, we propose a comprehensive timeline depicting the evolution of Transformer architectures in~\Cref{fig-timeline}.

\begin{table*}
    \caption{Comparison of different attention mechanism complexities. $N$: number of tokens/patches, $d$: embedding/channel dimension, $h$: number of heads, $hw$: window size, $M$: number of patches in a window, $HW$: image size, $r$: reduction ratio, $d_g$: number of groups, $K$: kernel size, $K_{DDW}$: deformable depth wise kernel size, $K_{DW}$: depth wise kernel size, $l$: level, $S_w$: stripe width, $H_0W_0$: coarsest level image size, $d_r$: dilation rate, $P$ is the concatenated sequence length of all pooled features}\label{tab-attention-complexity}
	\begin{center}
		\resizebox{\textwidth}{!}{
        \begin{tabular}{lrrrrccc}
			\bottomrule
            \rowcolor{gray!20}
			\textbf{Method} 
            & \textbf{Proposed at} 
            & \textbf{Params. (M)} 
            & \textbf{FLOPS (G)} 
            & \textbf{MACs (G)} 
            & \textbf{Computational Complexity ($O$)} 
            & \textbf{Link}
            \\ \hline

			\rowcolor{green!5}
            ViT \cite{dosovitskiy2020vit} 
            & 10-2020 
            & 8.40
            & 1.72 
            & 1.73
            & $N^{2}d$     
            & \href{https://github.com/google-research/vision_transformer}{link} 
            \\
            
            \rowcolor{green!10}
            Efficient Att. \cite{shen2021efficient} 
            & 11-2020 
            & 8.40
            & 1.69
            & 1.71
            & $d^2N$  
            & \href{https://github.com/cmsflash/efficient-attention}{link} 
            \\
			
            \rowcolor{green!5}
            XCiT \cite{ali2021xcit} ($p:16$)
            & 06-2021 
            & 12.62 
            & 2.52
            & 2.48   
            & $Nd^{2}/h$   
            & \href{https://github.com/facebookresearch/xcit}{link} 
            \\
            
            \rowcolor{green!10}
            P2T \cite{wu2022p2t}
            & 06-2021 
            & 12.65
            & 2.55 
            & 2.49
            & $(N + 2P)d^{2} + 2NPd$
            & \href{https://github.com/yuhuan-wu/P2T}{link} 
            \\
            
            \rowcolor{green!5}
            KVT \cite{wang2022kvt}
            & 06-2021 
            & 12.60 
            & 2.54
            & 2.48
            & - 
            & \href{https://github.com/damo-cv/KVT}{link} 
            \\
            
            \rowcolor{green!10}
            CSWin (tiny) \cite{dong2022cswin}
            & 07-2021 
            & 12.61 
            & 2.55 
            & 2.48
            & $Nd(4d + S_w H + S_w W)$ 
            & \href{https://github.com/microsoft/CSWin-Transformer}{link} 
            \\
            
            
            \rowcolor{green!10}
            QuadTree att. \cite{zhang2023quad} 
            & 01-2022 
            & 12.64
            & 2.55 
            & 2.49 
            & $2(H_0^2W_0^2+ 4/3(1 - 4^{1-l})KN)$  
            & \href{https://github.com/Tangshitao/QuadTreeAttention/tree/master}{link} 
            \\
            
            \rowcolor{green!5}
            MISSFormer \cite{huang2022missformer} 
            & 05-2022 
            & 12.66 
            & 3.38 
            & 3.31 
            & $\frac{2N^2d}{r^2} + Nd^2r^2$ 
            & \href{https://github.com/ZhifangDeng/MISSFormer}{link} 
            \\
            
            
            
            \rowcolor{green!5}
            Castling-ViT \cite{you2023castlingvit}
            & 11-2022 
            & 4.20 
            & 0.95
            & 0.83 
            & - 
            & \href{https://github.com/GATECH-EIC/Castling-ViT}{link} 
            \\
            
            
            \rowcolor{green!10}
            EfficientFormerV2 (l)\cite{li2023rethinking}
            & 12-2022 
            & 12.61 
            & 2.55 
            & 2.48 
            & - 
            & \href{https://github.com/snap-research/EfficientFormer}{link} 
            \\
            
            \rowcolor{green!5}
            DilateFormer (tiny) \cite{jiao2023dilateformer}
            & 02-2023 
            & 12.60 
            & 2.47 
            & 2.48 
            & - 
            & \href{https://github.com/JIAOJIAYUASD/dilateformer}{link} 
            \\
            
            \rowcolor{green!10}
            SwiftFormer \cite{shaker2023swiftformer} 
            & 03-2023 
            & 14.71 
            & 2.88 
            & 2.89
            & -
            & \href{https://github.com/Amshaker/SwiftFormer}{link} 
            \\
            
            
            \rowcolor{green!5}
            FLatten transformer \cite{han2023flatten} 
            & 08-2023 
            & 4.40
            & 0.87
            & 0.84 
            & $Nd^{2}$
            & \href{https://github.com/LeapLabTHU/FLatten-Transformer}{link} 
            \\
            
            \hline
            \rowcolor{green!10}
            LV-ViT \cite{jiang2021all} 
            & 04-2021 
            & 12.59 
            & 2.55 
            & 2.48 
            & - 
            & \href{https://github.com/zihangJiang/TokenLabeling}{link} 
            \\
            
            \rowcolor{green!10}
            MSG \cite{fang2022msg} 
            & 05-2021 
            & 12.59 
            & 10.23 
            & 0.31 
            & $12(HW)d^2+2HWd(hw)^2$ 
            & \href{https://github.com/hustvl/MSG-Transformer}{link} 
            \\
            
            \rowcolor{green!10}
            DynamicViT \cite{rao2021dynamicvit} 
            & 06-2021 
            & 12.59 
            & 2.55 
            & 2.48
            & - 
            & \href{https://github.com/raoyongming/DynamicViT}{link} 
            \\
            
            \rowcolor{green!5}
            CMT \cite{guo2022cmt}
            & 07-2021 
            & 12.66 
            & 2.57
            & 2.50 
            & $10Nd^2(1 + \frac{0.2}{k^2})+ \frac{2N^2d}{k^2}+ 45Nd$ 
            & \href{https://github.com/ggjy/CMT.pytorch}{link} 
            \\
            
            
            
            
            \rowcolor{green!10}
            MaxViT (b)\cite{tu2022maxvit}
            & 04-2022 
            & 28.89 
            & 5.72 
            & 5.57 
            & - 
            & \href{https://github.com/google-research/maxvit}{link} 
            \\
            
            
            \rowcolor{green!10}
            LITv2 (l)\cite{pan2023fast}
            & 05-2022 
            & 12.63 
            & 2.55 
            & 2.49 
            & $(\frac{7Nd^2}{4} + (hw)^2Nd)+((\frac{3}{4} + \frac{1}{(hw)^2}N^2d))$ 
            & \href{https://github.com/ziplab/LITv2}{link} 
            \\
            
            \rowcolor{green!5}
            HorNet (l)\cite{rao2022hornet}
            & 07-2022 
            & 12.35 
            & 2.42 
            & 2.43 
            & - 
            & \href{https://github.com/raoyongming/HorNet}{link} 
            \\
            
            
            
            \rowcolor{green!5}
            BiFormer \cite{zhu2023biformer} 
            & 03-2023 
            & 12.62 
            & 2.48
            & 1.87 
            & $3Nd^2 + 3dk^\frac23(2N)^\frac43$ 
            & \href{https://github.com/rayleizhu/BiFormer}{link} 
            \\
            
            
            \rowcolor{green!10}
            BViT (S)\cite{li2023bvit}
            & 06-2023 
            & 8.40 
            & 1.73 
            & 1.73 
            & - 
            & \href{https://github.com/koala719/BViT}{link} 
            \\
            
            \hline
            \rowcolor{green!5}
            PvT \cite{wang2021pyramid}
            & 02-2021 
            & 12.59 
            & 2.55 
            & 2.48 
            & $\frac{2N^2d}{r^2} + Nd^2r^2$ 
            & \href{https://github.com/whai362/PVT}{link} 
            \\
            
            \rowcolor{green!10}
            Swin Transformer \cite{liu2021swin} 
            & 03-2021 
            & 12.60
            & 2.49 
            & 2.48 
            & $4Nd^2 + 2(hw)^2Nd$ 
            & \href{https://github.com/microsoft/Swin-Transformer}{link} 
            \\
            
            
            
            
            
            \rowcolor{green!10}
            PoolFormer \cite{yu2022metaformer}
            & 11-2021 
            & 8.40 
            & 1.65 
            & 1.66 
            & - 
            & \href{https://github.com/sail-sg/poolformer}{link} 
            \\
            
            \rowcolor{green!5}
            Swin Transformer V2\cite{liu2021swinv2} 
            & 11-2021 
            & 12.60 
            & 2.49 
            & 1.86 
            & - 
            & \href{https://github.com/microsoft/Swin-Transformer}{link} 
            \\
            
            \rowcolor{green!10}
            GCViT \cite{hatamizadeh2023global} 
            & 06-2022 
            & 34.65 
            & 2.55 
            & 2.69 
            & $2HW(2d^2+hwd)$ 
            & \href{https://github.com/NVlabs/GCVit}{link} 
            \\
            
            \rowcolor{green!5}
            Slide-Transformer \cite{pan2023slide} 
            & 04-2023 
            & 9.45 
            & 1.87 
            & 1.88 
            & - 
            & \href{https://github.com/LeapLabTHU/DAT/tree/main/models}{link} 
            \\
            
            
            \rowcolor{green!5}
            FasterViT \cite{hatamizadeh2023fastervit} 
            & 04-2023 
            & 26.27 
            & 2.62 
            & 2.61 
            & $K^2H^2d + LH^2d + H^4/K^2 L^2d$ 
            & \href{https://github.com/NVlabs/FasterViT}{link} 
            \\
            
            
            \hline
            
            
            
            
            
            
            \hline
            
            \rowcolor{green!5}
            DeepViT \cite{zhou2021deepvit} 
            & 03-2021 
            & 12.59 
            & 2.55 
            & 2.48 
            & $2dN(M+d_g)$ 
            & \href{https://github.com/zhoudaquan/dvit_repo}{link} 
            \\
            
            \rowcolor{green!10}
            CvT \cite{wu2021cvt} 
            & 03-2021 
            & 12.63 
            & 2.55 
            & 2.49 
            & $K^2d$
            & \href{https://github.com/microsoft/CvT/tree/main}{link} 
            \\
            
            \rowcolor{green!5}
            LeViT \cite{graham2021levit} 
            & 04-2021 
            & 12.60 
            & 2.55 
            & 2.48 
            & - 
            & \href{https://github.com/facebookresearch/LeViT}{link} 
            \\
            
            \rowcolor{green!10}
            Deformable Att. \cite{xia2022vision} 
            & 01-2022 
            & 12.60 
            & 2.49 
            & 2.48 
            & $2\frac{(N)^2}{r^2} + 2Nd^2 + 2\frac{Nd^2}{r^2} + (k^2 + 2)\frac{HWd}{r^2}$ 
            & \href{https://github.com/LeapLabTHU/DAT}{link} 
            \\
            
            \rowcolor{green!5}
            FocalNet \cite{yang2022focal} 
            & 03-2022 
            & 12.60 
            & 2.49 
            & 2.49 
            & $N(3d^2 + d(2l+3)+d\sum_l(k^l)^2)$ 
            & \href{https://github.com/microsoft/FocalNet}{link} 
            \\
            
            \rowcolor{green!10}
            EViT \cite{feng2022evit} 
            & 08-2022 
            & 12.59
            & 2.55 
            & 2.48 
            & - 
            & \href{https://github.com/youweiliang/evit}{link} 
            \\
            
            
            
            
            
            \rowcolor{green!5}
            VAN \cite{Guo2023}
            & 06-2023 
            & 11.67 
            & 2.29 
            & 2.30 
            & $(K/d_r)^2d^2N$ 
            & \href{https://github.com/Visual-Attention-Network}{link} 
            \\
            
            \rowcolor{green!10}
            Bidirectional att. \cite{fan2023lightweight} 
            & 06-2023 
            & 14.92 
            & 2.17 
            & 2.17 
            & - 
            & \href{https://github.com/qhfan/FAT}{link} 
            \\
            
            \rowcolor{green!5}
            D-LKA  \cite{azad2023selfattention}
            & 11-2023 
            & 9.42 
            & 1.83
            & 1.86 
            & $d(d+2K_{DDW}^4+K_{DDW}^2d+2K_{DW}^4+ K_{dW}^2d)\times N$ 
            & \href{https://github.com/xmindflow/deformableLKA}{link} 
            \\
			
            \bottomrule
		\end{tabular}
	}
	\end{center}
\end{table*}

Guided by the primary objective of this paper, our focus remains squarely on improving the efficiency of ViTs. Recognizing the inherent reliance of transformer networks on attention mechanisms, particularly when contrasted with CNNs, due to their attention-centric structure, our exploration delves deeply into transformer architectures and a subset of hybrid transformer models. Hence it is readily apparent that attention mechanisms must have:

\begin{itemize}
	\item Portability and modularity across diverse networks.
	\item Maximal efficiency.
	\item Rich representation of the input.
\end{itemize}

Within the spectrum of proposed categories for enhancing ViTs, various architectures concentrate on optimizing tokenization, mitigating self-attention complexity, designing hierarchical feature representation~\cite{liu2021swin,liu2021swinv2}, utilizing channel and spatial attentions~\cite{ding2022davit,shaker2022unetr++} or incorporating alternative approaches to enhance overall performance. Strategies focused on rethinking tokenization aim to introduce additional tokens with supplementary information, reduce redundant tokens, or alter token meanings~\cite{fang2022msg,tu2022maxvit,zhu2023biformer}. Conversely, those targeting self-attention reduction strive to minimize the number of tokens, shift calculations to the channel dimension, and alter the order of query, key, and value operations~\cite{shen2021efficient,ali2021xcit,liu2023efficientvit}, albeit often at the expense of computational efficiency and model accuracy. Tailored ViTs leveraging alternative approaches, such as CVT~\cite{wu2021cvt}, Deformable attention~\cite{xia2022vision}, or Bidirectional interaction~\cite{fan2023lightweight}, incorporate convolution blocks to model local features, propose query-agnostic offsets for shifting keys and values to crucial regions and utilizing bidirectional interaction between local and global features. Despite these advancements, challenges persist, including computational resource constraints and the imposition of heavy-weight architectures.

Evidently, the categories of reducing self-attention, rethinking tokenization, and employing additional approaches have garnered substantial research attention, signaling a collective endeavor to develop efficient transformers. Notably, some proposed ViTs exhibit overlapping contributions; for instance, Biformer~\cite{zhu2023biformer}introduces a novel approach to utilizing context information for feature enhancement while concurrently incorporating a hierarchical design.

In summary, our paper aims to provide a detailed overview of the advancements in ViTs by organizing essential information in~\Cref{tab-transformer-architectures-1},~\Cref{tab-attention-complexity}, our taxonomy, and a timeline in~\Cref{fig-taxonomy-design} and~\Cref{fig-timeline}. This organizational framework is designed to offer the community a comprehensive and illuminating resource for understanding the evolution and improvements in ViTs.

\section{Further Analysis}\label{analysis-sec}

In this section, we conduct a comprehensive analysis of various ViT blocks, focusing on key factors such as issuance, number of parameters, FLOPs (Floating Point Operations), MACs (Multiply-Accumulate Operations), and computational complexity. Initially, experiments were performed using a conventional ViT network architecture~\cite{dosovitskiy2020vit} as a testing platform in~\Cref{tab-attention-complexity}. To evaluate ViT blocks based on our innovative classification, we modified the architecture by replacing the multi-head self-attention block with the specific blocks corresponding to each network category. Additionally, settings for the ViT main network were introduced to ensure a fair review. The input image size was set to 224 $\times$ 224, with 3 channels, a patch dimension of 16, 8 heads, an embedding dimension of 1024, an MLP layer dimension of 2048, and an increase rate of 4 for feed forward layer.

\Cref{tab-attention-complexity} is categorized into five sections, aligning with the introduced article categories: self-attention complexity reduction, rethinking tokenization approaches, hierarchical vision transformers, networks utilizing channel and spatial approaches, and ViT blocks employing various methods for performance enhancement. All experiments are compared against the basic ViT network in the first row.

Efficient Attention~\cite{shen2021efficient} and XCiT~\cite{ali2021xcit} effectively tackle the quadratic complexity challenge inherent in the vanilla Vision Transformer. They exhibit time complexities of $d^2N$ and $Nd^{2}/h$, respectively Flatten Transformer~\cite{han2023flatten}, with a time complexity of $Nd^{2}$, stands out for its parameter efficiency (4.40M parameters) compared to the vanilla ViT~\cite{dosovitskiy2020vit} (8.40M). These enhanced ViTs employ diverse techniques to mitigate quadratic complexity, providing options catering to specific computational efficiency and model size considerations.

Within the Rethinking Tokenization approaches, CMT~\cite{guo2022cmt}, LiT V2~\cite{pan2023fast}, and BiFormer~\cite{zhu2023biformer} exhibit variations in computational characteristics. CMT introduces a mix of linear and quadratic terms in its time complexity based on embedding dimension and kernel size, while LiT V2 combines linear and window-dependent quadratic terms. Notably, BiFormer stands out with a linear time complexity, suggesting a potential reduction in quadratic dependence on the number of tokens. BiFormer emerges as a promising option, potentially offering advantages in computational efficiency and model performance while utilizing a novel approach to computing attention at both the region and token levels.

Among the Hierarchical Vision Transformers, PvT~\cite{wang2021pyramid} and Swin~\cite{liu2021swin} demonstrate favorable characteristics with time complexities that do not exhibit quadratic dependency on the number of tokens $(N)$. PvT's time complexity includes a reduction in quadratic complexity of $N$, while Swin's complexity is linearly related to the number of tokens. Conversely, Hierarchical Attention~\cite{hatamizadeh2023fastervit} showcases a more FLOPs, MACs and number of parameters. In addressing the quadratic complexity challenge in vision transformers, PvT and Swin offer promising alternatives with reduced dependence on the number of tokens.

Within the category of ViTs employing various approaches, CvT~\cite{wu2021cvt} stands out for its linear dependency on the number of tokens. DLKA~\cite{azad2023selfattention} introduces an effective time complexity, potentially involving quadratic dependencies on Dilated and Depthwise Deformable Convolution Kernel size, which is much lower than the number of tokens in multihead self attention. Deformable Attention~\cite{xia2022vision} attempts to reduce the quadratic complexity of $N$ by using a reduction ratio but still suffers from quadratic terms. In addressing the quadratic complexity challenge in vision transformers, CvT and DLKA appear more favorable with their linear dependence on N, potentially offering advantages in terms of computational efficiency, number of parameters, and scalability.

In summary an optimal attention module should address the quadratic memory challenge while maintaining universality across various tasks. It should prioritize both speed and memory, emphasizing simplicity, avoiding rigid hard-coded elements or excessive engineering, and highlighting elegance and scalability. This comparative analysis serves as a valuable resource for understanding nuanced trade-offs among different ViT modules, aiding researchers and practitioners in selecting the most suitable architecture based on specific task requirements and constraints.
\section{Challenges and Future Aspects} 
\label{sec:challenge}



Despite the remarkable performance of ViT networks and their efficiency-enhanced counterparts, practical applications face several challenges. Key obstacles include the demand for substantial training data, the need for interpretability, real-time applicability, effective feature representation, and the associated high computational costs. In this section, we explore these challenges and outline future directions, aiming to provide researchers with valuable insights into the limitations and opportunities for developing more efficient versions of Transformer models. This investigation extends beyond ViT architectures, encompassing emerging paradigms such as Multi-Modal Transformers and Foundational models.

\subsection{Intensive Computational Requirements} 

The adaptability of Transformer models to high parametric complexity across various data modalities is a notable strength. However, this flexibility comes at a cost, as evidenced by the substantial training and inference costs associated with large-scale models. For instance, the basic BERT model~\cite{devlin2019bert}, with 109 million parameters, required approximately 1.89 peta-flop days for training, while the latest GPT-3 model~\cite{brown2020language}, with 175 billion parameters, demanded an astonishing 3640 peta-flop days for training~\cite{Khan_2022}. 

An empirical study on ViT networks scalability~\cite{dehghani2023scaling} indicates that scaling up in terms of compute, model size, and training samples improves performance. The study underscores that only large models, with more parameters, benefit from additional training data, while smaller models quickly reach a performance plateau and cannot leverage additional data. Although large-scale models possess the potential to enhance representation learning capabilities, their current designs are computationally prohibitive, necessitating the development of more efficient designs based on specific criteria~\cite{azad2023advances}.

The computational cost of Transformer models poses a significant challenge for CV applications. The time and memory cost of the core self-attention operation in Transformers increases quadratically with the number of input tokens $(O(n^{2}))$, where $n$ represents the number of image patches. Numerous proposed methods, discussed in \Cref{chap:attention-design}, aim to make ViTs more \textit{'efficient'} by employing strategies such as Self Attention Complexity Reduction~\cite{shen2021efficient,ali2021xcit,shaker2023swiftformer}, Rethinking Tokenization~\cite{fang2022msg,rao2022hornet,zhu2023biformer}, Channel and Spatial Transformers~\cite{ding2022davit,sun2022fusing,shaker2022unetr++}, Hierarchical Vision Transformers~\cite{wang2022pvt,cao2022swin,hatamizadeh2023fastervit}, and other approaches~\cite{yang2022focal,fan2023lightweight,azad2023selfattention} categorized based on their design choices. However, most of these approaches involve a trade-off between complexity and accuracy, necessitate specialized hardware, or are limited in applicability to high resolution images. Consequently, there is an urgent need to develop ViT models with enhanced attention mechanisms designed for various CV tasks, suitable for resource-limited systems without compromising accuracy. Exploring how existing models can reduce computational costs adds an interesting dimension to this challenge.

\subsection{Extensive Data Demands and Feature Representation}
As ViT architectures lack inherent inductive biases specialized to visual data, they often demand extensive training to decipher modality-specific rules. Unlike CNNs, which incorporate built-in features such as translation invariance, weight sharing, and partial scale invariance, ViTs must autonomously deduce these image-specific concepts from training examples~\cite{wu2021cvt,dai2021coatnet}. This necessity leads to prolonged training durations, a substantial increase in computational requirements, and a reliance on extensive datasets. For instance, the ViT~\cite{dosovitskiy2020vit} model requires training on hundreds of millions of image examples to achieve satisfactory performance on the ImageNet benchmark dataset.

Efforts have been made to address this challenge. For instance, the DeiT~\cite{touvron2021training} model adopts a distillation approach to enhance data efficiency. Moreover, integrating CNN-like feature hierarchies~\cite{wang2022pvt, liu2021swin, liu2021swinv2, chen2021regionvit} or directly embedding stacked Convolutional blocks within the ViT architecture~\cite{wu2021cvt,xu2021levit,hatamizadeh2023fastervit,dai2021coatnet} provides an opportunity for modified ViT models to be trained effectively on smaller datasets and capture local and global features simultaneously. This approach introduces flexibility and offering promising avenues for future developments in ViT efficiency.

\subsection{Multi-Modal Transformers and Foundational Models}

Leveraging the intrinsic advantages and scalability of transformers in modeling diverse modalities and tasks, such as language, visual, and auditory inputs, has sparked interest in the development of Multi-Modal Transformers (MMTs)~\cite{xu2023multimodal}. Unlike traditional models burdened by modality-specific architectural assumptions, MMTs showcase flexibility by accommodating one or multiple sequences of tokens as input. This inclusivity allows for seamless integration of Multi-Modal Learning (MML) without the need for extensive architectural modifications~\cite{jaegle2021perceiver}. The simplicity of learning per-modal specificity and inter-modal correlation is achieved by manipulating the input pattern of self-attention. The surge in research endeavors across disciplines has resulted in the emergence of numerous novel MML methods in recent years, contributing significantly to advancements in various domains~\cite{devlin2019bert,dosovitskiy2020vit,carion2020end,sun2019videobert,NEURIPS2021_344ef515}.

Simultaneously, a parallel trend in the exploration of large foundation models (LFMs) has emerged, akin to Language Models (LLMs) in NLP. Notably, pre-trained Vision-Language Models (VLM)~\cite{azad2023foundational}, exemplified by SAM~\cite{kirillov2023segment}, exhibit promising zero-shot performance in diverse vision tasks like class-agnostic segmentation given an image and a visual prompt such as a box, point, or mask. SAM is trained on billions of object masks following a model-in-the-loop (semi-automated) dataset annotation setting. Such generic visual prompt-based segmentation models can be adapted for specific downstream tasks, including medical image segmentation~\cite{ma2023segment}, robotics~\cite{yang2023transferring}, and real-time vision~\cite{chen2023rsprompter}.

While Multi-Modal Transformers and foundational models are distinct concepts, there exists an intriguing overlap in their exploration. The former emphasizes the synergy of diverse modalities within a unified transformer framework, while the latter delves into the development of large foundation models, such as VL models and visual prompt-based segmentation models, for various perception tasks. This dynamic landscape highlights the evolving nature of transformer-based architectures in accommodating and enhancing multi-modal learning and foundational model development. Notably, the challenges associated with developing these models include addressing modality-specific intricacies, managing data heterogeneity, and optimizing computational efficiency, all of which contribute to the complexity of pushing the boundaries in transformer-based model design. Undoubtedly, this dynamic field provides fertile ground for future research endeavors.

\subsection{Explainability}

With the recent advancements in Explainable Artificial Intelligence (XAI) and the development of algorithms aiming to enhance interpretability in Deep Learning, researchers are actively working on integrating XAI methods into the construction of transformer-based models. This effort seeks to establish more reliable and comprehensible systems across various domains, including applications in CV tasks~\cite{ali2022xai,azad2023advances,Komorowski_2023}. Despite the robust performance of transformer architectures, there is a growing need to unravel the decision-making processes within these models. This involves visualizing pertinent regions in an image that influence a specific classification decision~\cite{qiang2023interpretabilityaware}. ViTs offer the unique capability of generating attention maps that highlight correlations between input regions and predictions.

However, a notable challenge arises as attention from each layer becomes intricately intertwined in subsequent layers, creating a complex structure that complicates the visualization of the relative contribution of input tokens toward final predictions~\cite{abnar2020quantifying}. This intricacy poses a significant hurdle in understanding the decision-making mechanism of ViTs. Additionally, challenges such as numerical instabilities in propagation-based XAI methods like LRP~\cite{LRP} and the inherent vagueness of attention maps, leading to inaccurate token associations~\cite{pmlr-v162-kim22g}, underscore the need for further research to enhance the interpretability of ViTs in the field of CV. This ongoing exploration represents an open research opportunity to unravel the nuances of interpretability in ViT networks.

\subsection{Real-Time Applicability}

In the pursuit of advancing the efficiency of ViT models, a critical consideration lies in their real-time applicability, particularly in resource-constrained environments like mobile devices. The integration of these models into such settings not only extends advanced vision capabilities to a broader user base but also aligns with the growing emphasis on eco-friendly practices within AI~\cite{chen2023pixartalpha}. This adaptability to constrained environments contributes to lowering deployment costs and fosters a more sustainable approach in model development.

Addressing challenges in real-time mobile vision tasks has become a focal point for researchers. The limitations of self-attention in real-time applications, especially on resource-constrained mobile devices, have led to the exploration of hybrid approaches that balance the advantages of convolutions and self-attention~\cite{mehta2022mobilevit}. Despite these efforts, the bottleneck of expensive matrix multiplication operations in self-attention persists, necessitating the development of more efficient models~\cite{shaker2023swiftformer,li2023rethinking}. This involves a strategic combination of CNNs and transformers, a critical consideration for mobile devices with limited computational resources~\cite{mehta2022mobilevit,hatamizadeh2023fastervit}.

In the domain of real-world applications demanding timely visual recognition on resource-constrained mobile devices, the imperative is to design lightweight and fast ViT models. Unlike their counterparts, lightweight CNNs, ViT-based networks face challenges in terms of optimization difficulties, extensive data augmentation requirements, and the need for expensive decoders~\cite{dosovitskiy2020vit, lepikhin2020gshard,wang2022pvt,xiao2021early}. Hybrid approaches that integrate convolutions and transformers are gaining traction, yet they often fall short of achieving true light-weight status~\cite{wu2021cvt, xu2022evo,chen2022mobileformer}. The quest to combine the strengths of CNNs and transformers for building robust and efficient ViT models for mobile vision tasks remains an open question. The emergence of solutions like MobileViT~\cite{mehta2022mobilevit}, FastViT~\cite{pan2023fast} and SwiftFormer~\cite{shaker2023swiftformer}, which address issues such as efficient additive attention and reduced computational complexity, highlights the evolving landscape in this dynamic field.

\section{Conclusion}

This paper surveyed existing literature focused on optimizing ViT models, particularly emphasizing the complexity associated with the self-attention module.
We outlined a taxonomy and high-level abstraction of the fundamental methods used in these new model classes and offered an extensive overview of various efficient transformer models. Besides, we discussed the landscape of these models, providing a detailed description of their design trends and the complexities of each block using comparison tables to highlight network parameters, FLOPS and other factors. We wrap up this survey by pinpointing research trends and future directions.

    

\appendix

\bibliographystyle{CVMbib}
\bibliography{reference.bib}


\end{document}